\documentclass[12pt]{article}

\usepackage{amsmath, amssymb}
\usepackage{booktabs, threeparttable, multirow}
\usepackage{graphicx}
\usepackage[round]{natbib}
\usepackage[colorlinks, citecolor=red]{hyperref}
\usepackage{authblk}
\usepackage{geometry,setspace}
\usepackage{pdfpages}

\def\top{{\mathrm{\scriptscriptstyle T} }}
\newcommand\independent{\protect\mathpalette{\protect\independenT}{\perp}}
\def\independenT#1#2{\mathrel{\rlap{$#1#2$}\mkern2mu{#1#2}}}
\DeclareMathOperator{\var}{var}

\DeclareMathOperator{\expit}{expit}

\DeclareMathOperator{\pr}{pr}
\DeclareMathOperator{\T}{T}
\DeclareMathOperator{\C}{C}
\allowdisplaybreaks[4]
\newtheorem{theorem}{Theorem}
\newtheorem{assumption}{Assumption}

\newtheorem{remark}{Remark}

\title{\bf Causal Inference with Truncation-by-Death and Unmeasured Confounding}

\author[1]{Yuhao Deng}
\author[2]{Yingjun Chang}
\author[1]{Xiao-Hua Zhou}
\affil[1]{\small ~Peking University}
\affil[2]{\small ~Peking University People's Hospital}
\date{}

\begin{document}

\maketitle

\begin{abstract}
Clinical studies sometimes encounter truncation by death, rendering outcomes undefined. Statistical analysis based solely on observed survivors may give biased results because the characteristics of survivors differ between treatment groups. By principal stratification, the survivor average causal effect was proposed as a causal estimand defined in always-survivors. However, this estimand is not identifiable when there is unmeasured confounding between the treatment assignment and survival or outcome process. In this paper, we consider the comparison between an aggressive treatment and a conservative treatment with monotonicity on survival. First, we show that the survivor average causal effect on the conservative treatment is identifiable based on a substitutional variable under appropriate assumptions, even when the treatment assignment is not ignorable. Next, we propose an augmented inverse probability weighting (AIPW) type estimator for this estimand with double robustness. Finally, large sample properties of this estimator are established. The proposed method is applied to investigate the effect of allogeneic stem cell transplantation types on leukemia relapse. \par
\end{abstract}

\section{Introduction}\label{sec1}

In clinical studies with mortality, outcomes other than mortality (such as quality of life following surgery) may be of primary interest. Since some individuals die before their primary outcomes are evaluated, we cannot measure those outcomes. This phenomenon is known as the truncation-by-death problem \citep{rubin2000discussion, mcconnell2008truncation}. Truncation-by-death is a different concept from censoring in that the latter indicates that an outcome exists but is not recorded, resulting in missing data, whereas the former renders the outcomes undefined. Because the observed survivors may consist of units with possibly different unmeasured characteristics, the observed outcomes in survivors between different treatment groups are not comparable.

A proper way to handle the truncation-by-death problem is principal stratification \citep{frangakis2002principal, rubin2006causal}. Principal strata are typically defined by the joint potential values of a post-treatment intermediate variable \citep{jiang2021identification}. Because the potential values of the intermediate variable for a unit are irrelevant to the treatment that the unit actually received, the principal stratum reflects the unit's intrinsic characteristics and acts as a baseline covariate. A meaningful causal contrast can only be defined in a subpopulation whose both potential outcomes are well defined, and such an estimand is often referred to as the survivor average causal effect (SACE). However, since the always-survivors are not individually observable, the interpretation of SACE should be paid extraordinary caution. 

To study the SACE, \citet{zhang2003estimation}, \citet{imai2008sharp}, \citet{chiba2011simple}, \citet{chiba2011large}, and \citet{long2013sharpening} provided bounds for this estimand. \citet{yang2016using} also used post-treatment information to construct bounds. However, these bounds are often too wide and not very useful for statistical inference. To point-identify the SACE, a commonly adopted assumption is the monotonicity of survival with respect to treatment. In addition to monotonicity, the relationship between the survival and outcome is sometimes assumed to follow a biased selection model or pattern mixture model with hyperparameters \citep{gilbert2003sensitivity, shepherd2006sensitivity, jemiai2007semiparametric}. In some studies, the biased selection model has been replaced with an independence assumption between the principal strata and potential outcomes in observed survivors, which is referred to as explainable nonrandom survival or principal ignorability \citep{hayden2005estimator, egleston2007causal, ding2017principal}. Sensitivity analysis can be performed to assess the feasibility of the biased selection model or independence assumption. 

If various types of covariates are available, information regarding survival can be leveraged to point-identify the SACE, which calls for a covariate playing a functioning role. Conditional independence is constructed using this key covariate. \citet{ding2011identifiability} considered the identification of the SACE based on a baseline covariate whose distribution is informative of the principal strata in randomized experiments, but assumed no common causes for the survival and outcome processes. A simulation study conducted by \citet{mcguinness2019comparison} indicated that unmeasured survival--outcome confounding would bias the estimation of the SACE. \citet{tchetgen2014identification} used post-treatment correlates of survival and outcomes in longitudinal studies to identify the SACE under nonparametric structural equations models. Further, \citet{wang2017identification} used a substitutional variable correlated with principal strata in addition to a variety of covariates to identify the SACE. 


However, most methods for handling the truncation-by-death problem assumed no unmeasured confounding between the treatment assignment and survival process.
These methods rely on ignorability so that the proportions of principal strata in both treatment groups are identical. Theoretically, statistical analysis becomes more challenging without ignorability because we would have no information about the distribution of principal strata. There are two possible ways to acheive identifiability in the presence of unmeasured confounding. The first is to restrict the relationship between the survival (principal strata) and outcome process (potential outcomes), such as introducing the explainable nonrandom survival assumption \citep{egleston2009estimation}, which omits the influence of principal strata on potential outcomes. The second way is to explore the unmeasured confoundedness between the treatment assignment and survival process by introducing a latent variable \citep{kedagni2021identifying}. However, this approach cannot identify the SACE because the proportions of principal strata are inestimable, unless there is more information about the outcome distribution in different principal strata. 

In this paper, we consider observational studies with unmeasured confounding when some outcomes are truncated by death. The unmeasured confounding matters in the following two aspects. On the one hand, the unmeasured confounding between the treatment--survival processes renders the distributions of principal strata different between treatment groups. The target population is hard to define and identify because the always-survivors in different treatment groups are different in unmeasured features. On the other hand, the unmeasured confounding between the survival--outcome processes results in a cross-world reliance of potential outcomes on potential survivals. The causal effect in always-survivors is hard to identify because the observed outcomes are mixed from more than one principal stratum. This setup is referred to as latent ignorability, where the principal strata (joint potential survivals) confound the treatment--survival--outcome processes simultaneously.

The innovation is twofold. First, we contribute to causal inference with unmeasured confounding by proposing a substitutional variable strategy. The substitutional variable is a baseline proxy of the always-survivorship (latent indicator of the target population) which do not require the equivalence or completeness assumption as required in auxiliary approaches or negative controls \citep{tchetgen2020introduction, miao2022identifying}. We show that the survivor average causal effect on the conservative treatment group (SACEC) can be identified with a single substitutional variable under certain assumptions, even though the SACE in the overall population is not identifiable. The core assumption on the substitutional variable can be assessed by sensitivity analysis.  Second, an augmented inverse probability weighting (AIPW) type estimator is proposed to estimate the SACEC with double robustness. Large sample properties of this AIPW type estimator are established.

The remainder of this paper is organized as follows. In Section \ref{sec2}, we introduce the motivating data, notations and causal estimands under the potential outcome framework. In Section \ref{sec3}, we present assumptions for identifying causal effects. Estimation and a sensitivity analysis framework are also illustrated in this section. Simulation studies are presented in Section \ref{sec6} to compare the proposed method to existing methods. The proposed method is applied to a retrospective allogeneic stem cell transplantation dataset to investigate the effects of transplantation types on leukemia relapse in Section \ref{sec7}. This paper concludes with a brief summary of our findings. The supporting information includes causal graphs, proofs of theorems and additional simulation results.

\section{Motiviting data and causal estimands}\label{sec2}

\subsection{Motivating data}

Allogeneic stem cell transplantation is a widely adopted approach to treat acute lymphoblastic leukemia, including human leukocyte antigen (HLA)-matched sibling donor transplantation (MSDT) and haploidentical stem cell transplantation (haplo-SCT) from family. The non-relapse mortality (NRM) for MSDT is lower than that for haplo-SCT becasue there are fewer mismatched HLA loci between the donor and patient undergoing MSDT, so doctors usually prefer MSDT.
In recent years, some benefits of haplo-SCT have been noted, in that stronger graft-versus-leukemia effects with haplo-SCT are found \citep{chang2017haploidentical, chang2020haploidentical}.

It is important to know whether haplo-SCT can achieve competitive performances with MSDT in terms of relapse if patients can survive, as the maturing operations are leading to higher survival rate. Formally speaking, we are interested in the treatment effect of transplantation types on relapse in always-survivors undergoing haplo-SCT. Statistical findings based on causal inference would help establish clinical guidelines for allogeneic stem cell transplantation.

Data from a retrospective study were collected at Peking University People's Hospital in China, including 1161 patients undergoing allogeneic stem cell transplantation from 2009 to 2017 \citep{ma2021an}. In the dataset, 21.53\% patients received MSDT (with no mismatched HLA loci) and 78.47\% received haplo-SCT (with mismatched HLA loci). By the end of two years after transplantation, the survival rate is 89.60\% for MSDT and 84.96\% for haplo-SCT. Among survivors, the relapse rate is 21.88\% for MSDT and 18.99\% for haplo-SCT.

\subsection{Estimands}

Consider an observational study with binary treatments such that $Z=1$ for the aggressive treatment (MSDT) and $Z=0$ for the conservative treatment (haplo-SCT). Let $S(z)$ be the binary potential survival ($z=0,1$), where $S(z)=1$ if the subject would survive and $S(z)=0$ otherwise. A potential outcome (incidence of relapse) $Y(z)$ is defined only if $S(z)=1$. We can supplementarily denote $Y(z)=*$ if $S(z)=0$. We assume a stable unit treatment value and consistency. Then $S=ZS(1)+(1-Z)S(0)$ and $Y=ZY(1)+(1-Z)Y(0)$ whenever an outcome is well defined. Baseline covariates $(V, X)$ on supports $\mathcal{V}\times\mathcal{X}$ are collected. Here $X$ act as measured confounders and $V$ acts as a substitutional variable for $S(0)$; a rigourous statement is deferred to next section. Let the subscript $i=1,\ldots,n$ denote the indexes of subjects. The observed sample $\{O_i=(Z_i, S_i, Y_i, X_i, V_i): i=1,\ldots,n\}$ is independently drawn from an infinite super-population.

Stratified by the joint values of $G=(S(1),S(0))$, there are four principal strata \citep{ding2011identifiability}: LL (always-survivors), LD (protected), DL (harmed), and DD (doomed); see Table \ref{tab:ps}. Often the DL stratum is excluded by a monotonicity assumption. The observed survivors in the aggressive treatment group come from the LL and LD strata, whereas the survivors in the conservative treatment group come from the LL stratum.

\begin{table}[h]
\centering
\caption{Principal strata (\checkmark means well defined and $*$ means undefined)} \label{tab:ps}
\begin{tabular}{ccccccc}
\toprule
Stratum & $S(1)$ & $S(0)$ & $Y(1)$ & $Y(0)$ & Description & \\ \midrule
LL & 1 & 1 & \checkmark & \checkmark & Always-survivors & \\
LD & 1 & 0 & \checkmark & $*$ & Protected & \\
DL & 0 & 1 & $*$ & \checkmark  & Harmed & (Excluded by monotonicity) \\
DD & 0 & 0 & $*$ & $*$ & Doomed & \\
\bottomrule
\end{tabular}
\end{table}

To define a meaningful causal estimand, we should restrict our analysis on the LL stratum, where both potential outcomes are well defined. The survivor average causal effect on the overall population (SACE), on the aggressive treatment (SACET) and on the conservative treatment (SACEC) are defiend as
\begin{align*}
&\Delta = E\{Y(1)-Y(0) \mid G=\text{LL}\}, \ \Delta_{\T} = E\{Y(1)-Y(0) \mid G=\text{LL}, Z=1\}, \\ &\Delta_{\C} = E\{Y(1)-Y(0) \mid G=\text{LL}, Z=0\},
\end{align*}
respectively. These three estimands generally are not equal because $Z$ is confounded with $G$. In some circumstances, the aggressive treatment may induce unsatisfactory quality of life despite survival. Practitioners would like to know whether the conservative treatment is sufficient for survivable patients, so they care about $\Delta_{\C}$. Another advantage of adopting $\Delta_{\C}$ as the estimand is that its target population is individually observed as the survivors receiving the conservative treatment.

\section{Identification and Estimation}\label{sec3}

\subsection{Identification of the causal estimand using a substitutional variable}

Suppose there is an unmeasured confounder $U$ between the treatment $Z$ and principal stratum $G$ \citep{kedagni2021identifying}. The backdoor paths from $Z$ to $(Y(1),Y(0))$ are blocked by $(V,X)$ and $G$. So we assume latent ignorability as follows, which is weaker than and can be implied by ignorability.

\begin{assumption}[Latent ignorability] \label{LI}
$Y(z) \independent Z \mid G=\text{LL}, V, X$ for $z=0,1$.
\end{assumption}

By controlling as many predictors of $Z$ and risk factors of $Y$ as possible, latent ignorability can be plausible.

\begin{assumption}[Monotonicity] \label{MO}
$S(1) \ge S(0)$ almost surely.
\end{assumption}
\begin{assumption}[Positivity] \label{PO}
$0<\pr(Z=1\mid V, X)<1$, $0<\pr(S(0)=1 \mid Z=1, V, X)<1$, and  $\pr(S(0)=1 \mid Z=0, V, X)>0$.
\end{assumption}

Monotonicity is often reasonable in observational studies. The aggressive treatment should produce a higher survival rate than the conservative treatment in a short period. Positivity and monotonicity ensure that always-survivors and doomed individuals exist in the aggressive treatment group.
To disentangle the mixture of LL and LD in this aggressive treatment group, a substitutional variable for $S(0)$ is needed.

\begin{assumption}[Substitution relevance] \label{SR}
$V \not\independent S(0) \mid Z=1, S(1)=1, X$.
\end{assumption}
\begin{assumption}[Nondifferential substitution] \label{NS}
$V \independent S(1) \mid Z=1, S(0)=0, X$.
\end{assumption}
\begin{assumption}[Non-interaction] \label{ER}
$E\{Y(z) \mid G, V, X\} = E\{Y(z) \mid G, X\} + f(V, X)$ for some maybe unknown function $f(V, X)$.
\end{assumption}

Substitution relevance helps distinguish LL and LD strata among observed survivors using the substitutional variable. Nondifferential substitution states that $V$ is a baseline proxy of $S(0)$, whose measurement error is irrelevant to $S(1)$. The substitutional variable can also be understood as a proxy of the always-survivorship in the $Z=1$ group, which contains information of the indicator of target population but does not contain irrelevant information. Nondifferential substitution can be implied by the combination of $V \independent S(1) \mid S(0)=0,X$ and $S(1) \independent Z \mid S(0)=0,V,X$. Non-interaction means that the effect of $V$ on $Y$ does not modify the effects of $G$ or $Z$ on $Y$. A special case of non-interaction is exclusion restriction, where $V$ does not have direct effect on $(Y(1),Y(0))$ so that $f(v, x)\equiv0$.

\begin{figure}
\centering
\includegraphics[width=0.4\textwidth]{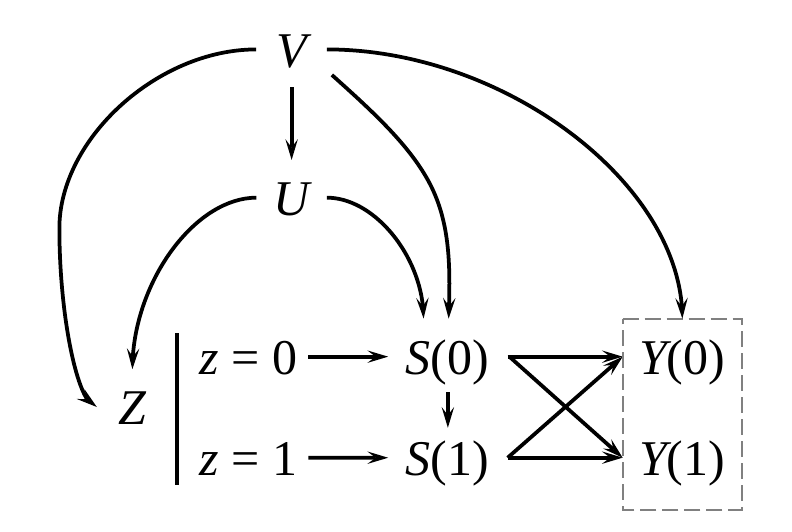}
\caption{A causal structural model. Baseline covariates $X$, which can have edges to $V$, $Z$, $U$, $S(0)$, $S(1)$, $Y(0)$ and $Y(1)$, are omitted for simplicity.} 
\label{swig}
\end{figure}

Figure \ref{swig} presents a possible causal structural model with an unmeasured confounder $U$ between $Z$ and $G=(S(1),S(0))$. Potential outcomes $(Y(1),Y(0))$ can depend on $G$. The substitutional variable has association with $S(0)$, but its association with $S(1)$ is blocked by $S(0)$ as a result of substitution relevance and nondifferential substitution. In Supplementary Material A, we give some examples that satisfy these six assumptions.

Define the propensity score, survival score and outcome regression model as
\begin{align*}
e(v,x) &= \pr(Z=1 \mid V=v, X=x), \\
\pi_{z}(v,x) &= \pr(S=1 \mid Z=z,  V=v, X=x), \\
m_{z}(v,x) &= E(Y\mid Z=z, S=1, V=v, X=x),
\end{align*}
respectively, for $z=0,1$. We denote the proportions of principal strata in the aggressive treatment group as $\pi_g(v,x)=\pr(G=g \mid Z=1, V=v, X=x)$ for $g \in \{\text{LL}, \text{LD}, \text{DD}\}$.

Denote $E_{V|X}\{g(V,X)\}=E\{g(V,X)\mid X\}$ for any measurable function $g(V,X)$ with finite expectation. Let $h(v,x)$ be an integrable function with respect to $v$ subject to 
$$E_{V|X}\left\{h(V,X)\right\} = 0, \quad E_{V|X}\left\{\frac{\pi_{\text{LL}}(V,X)}{\pi_{\text{LD}}(V,X)}h(V,X)\right\} \neq 0.$$
Define weight functions
\begin{align*}
R(v,x) &= \frac{\frac{\pi_1(v,x)}{1-\pi_1(v,x)}h(v,x)}{E_{V|X=x}\left\{\frac{\pi_1(V,X)}{1-\pi_1(V,X)}h(V,X)\right\}}, \\
R_0(v,x) &= \frac{(1-e(v,x))\pi_0(v,x)}{\pr(Z=0, S=1)}, \quad R_1(v,x) = R(v,x) E_{V|X=x}\{R_0(V,X)\}.
\end{align*}
The function $R_0(v,x)$ represents a covariate shift from the overall population to the survivors (always-survivors) in the conservative treatment group ($Z=0$). We have the following identification result.

\begin{theorem} \label{thm2}
Under Assumptions 1--6, SACEC can be identified,
\begin{align}
\Delta_{\C} &= E\left[R_1(V,X)\{m_1(V,X)-m_0(V,X)\}\right] \label{SACEC1} \\
&= E\left[R_1(V,X)\left\{\frac{ZS(Y-m_1(V,X))}{e(V,X)\pi_1(V,X)}+m_1(V,X)-m_0(V,X)\right\}\right. \notag \\
&\qquad\qquad\qquad \left. - R_0(V,X)\left\{\frac{(1-Z)S(Y-m_0(V,X))}{(1-e(V,X))\pi_0(V,X)}\right\}\right] \label{SACEC2}.
\end{align}
\end{theorem}

\begin{remark}
If $V$ is not binary, then there are many possible ways to construct $h(v,x)$ since only two values of $V$ are required for identification. Different choices of $h(v,x)$ can lead to a unique solution of $\Delta_{\C}$. For example, we can take $h(v,x) = v - E(V|X=x)$.
\end{remark}

The proof of Theorem \ref{thm2} is given in Supplementary Material B. The proportion of always-survivors in the $Z=1$ group is not identifiable under Assumptions 1--6, so SACE and SACET are not identifiable. 

Now we give three examples to illustrate when this identification strategy based on the substitutional variable can be reasonable.

\textit{Example 1}.
Suppose a traditional therapy (such as chemotherapy) has been widely adopted to cure a disease, but this therapy may cause adverse effects (readmission). Now a novel therapy (such as medication) is invented, which can greatly lower the safety risk, so monotonicity holds. In a cohort study, patients decided proper treatments to take based on doctor's suggestion. Let $Z$ be the treatment a patient received, $S$ be the occurence of readmission, and $Y$ be the outcome of interest, such as quality of life. We want to measure the treatment effect on quality of life at home. The substitutional variable $V$ can be a baseline risk factor of the adverse effect for the chemotherapy. By definition, $V$ satisfies substitution relevance and nondifferential substitution since $V$ is irrelevant to medication outcomes. Suppose the risk factor $V$ would not affect the quality of life if this patient can live normally outside the hospital, so exclusion restriction (non-interaction) is also satisfied.

\textit{Example 2}.
Suppose a targeted durg is used to cure cancer. This targeted drug is effective only if the patient carry certain gene. Now a generic drug targeting on a wider range of but unconfirmed population is invented. Let $Z$ indicate which drug to take, $S$ be an indicator indicating whether a patient response to the drug, and $Y$ be the outcome of interest. Practitioners want to know whether the generic drug has a better performance than the targeted drug. To make these two drugs comparative, the study should condition on the subpopulation who are responsive to both drugs. In a clinical trial, patients with PCR testing positive for the targeted gene are randomized to take the targeted drug or generic drug, while patients with PCR testing negative are all assigned to take the generic drug. Monotonicity means that the patients with PCR testing positive are also eligible for the generic drug, ensured by the wider target of the generic drug. The substitutional variable $V$ can be the PCR test result. It satisfies substitution relevance and nondifferential substitution because the PCR test cannot predict the response to the generic drug if a patient does not respond to the targeted drug. It also satisfies non-interaction because PCR test results cannot directly influence outcomes: it is $G=(S(1),S(0))$ that may influence outcomes.

\textit{Example 3}.
Suppose the government puts forward a training program to promote employment for job finders. Here $Z$ is whether a person joins in the program, $S$ is an indicator of employment, and $Y$ is the earning if employed. The government wants to know the effect of this program on earnings, so the study should restrict to the always-employed subpopulation. A person who tends to be unemployed would be more willing to join in the program. Monotonicity means that joining in the program would make job finding easier. The substitutional variable $V$ can be self-reported confidence of successfully finding a job before joining in the program, so $V$ satisfies substitution relevance and nondifferential substitution. Such a variable $V$ can be collected by questionnaire when signing up for the program. Non-interaction is also reasonable because once he/she is employed, the pre-treatment confidence for being employed measured before joining in the training program would not make a difference on earnings whether he/she joined in the training program or not.

\subsection{Estimation of the causal effect}

Suppose $e(v,x)$, $\pi_z(v,x)$, $m_z(v,x)$ and $R_z(v,x)$ are estimated as $\widehat{e}(v,x)$, $\widehat{\pi}_z(v,x)$, $\widehat{m}_z(v,x)$ and $\widehat{R}_z(v,x)$, respectively for $z=0,1$. 
The denominator of $R(v,x)$ can be estimated by empirical average in a neighborhood of each $(v,x)$ based on the Mahalanobis distance or kernel.

We focus on the property of the model-assisted AIPW type estimator
\begin{align}
\widehat\Delta_{\C} &= \mathbb{E}_n \left[\widehat{R}_1(V,X)\left\{\frac{ZS(Y-\widehat{m}_1(V,X))}{\widehat{e}(V,X)\widehat{\pi}_1(V,X)}+\widehat{m}_1(V,X)-\widehat{m}_0(V,X)\right\}\right. \notag \\
&\qquad\qquad\qquad \left. - \widehat{R}_0(V,X)\left\{\frac{(1-Z)S(Y-\widehat{m}_0(V,X))}{(1-\widehat{e}(V,X))\widehat{\pi}_0(V,X)}\right\}\right], \label{AIPWe}
\end{align}
where $\mathbb{E}_n(\cdot)$ means empirical average on the full sample. Let $\widehat\Delta_{\C}^{*}$ be the oracle estimator in which all models are known,
\begin{align}
\widehat\Delta_{\C}^{*} = \mathbb{E}_n\{\phi(O)\} &= \mathbb{E}_n \left[R_1(V,X)\left\{\frac{ZS(Y-m_1(V,X))}{e(V,X)\pi_1(V,X)}+m_1(V,X)-m_0(V,X)\right\}\right. \notag \\
&\qquad\qquad\qquad \left. - R_0(V,X)\left\{\frac{(1-Z)S(Y-m_0(V,X))}{(1-e(V,X))\pi_0(V,X)}\right\}\right]. \label{AIPWo}
\end{align}
The oracle estimator $\widehat\Delta_{\C}^{*}$ enjoys asymptotic normality $n^{1/2}(\widehat\Delta_{\C}^{*}-\Delta_{\C}) \rightarrow_{d} N(0, C)$, where 
\[
C = E\left\{\frac{R_1(V,X)^2\var(Y|V,X,Z=1)^2}{e(V,X)\pi_1(V,X)}+\frac{R_0(V,X)^2\var(Y|V,X,Z=0)^2}{(1-e(V,X))\pi_0(V,X)}\right\} + \var\{\Delta(X)\}.
\]
There is higher-order bias between $\widehat\Delta_{\C}$ and $\widehat\Delta_{\C}^{*}$ due to the variation of estimated models. By cross-fitting \citep{chernozhukov2018double}, the discrepency can be described by a drift term converging to zero in probability under correct model specifications,
\begin{align*}
D_n &= \mathbb{E}_n \Big\{(\widehat{R}_1(V,X)-R_1(V,X))(m_1(V,X)-m_0(V,X)) \\
&\qquad\qquad\qquad -(R_1(V,X)-R_0(V,X))(\widehat{m}_0(V,X)-m_0(V,X))\Big\}.
\end{align*}
The large sample properties of $\widehat\Delta_{\C}$ is demonstrated in the following theorem.

\begin{theorem} \label{thm3}
(a) If $\{\widehat{m}_0(v,x), \widehat{R}_z(v,x): z=0,1\}$ are sup-norm consistent upon $(v,x)\in\mathcal{V}\times\mathcal{X}$, then $\widehat\Delta_{\C}^{*}$ and $\widehat\Delta_{\C}$ converge in probability to $\Delta_{\C}$ if either $\{\widehat{e}(v,x), \widehat{\pi}_z(v,x): z=0,1\}$ or $\widehat{m}_1(v,x)$ is sup-norm consistent upon $(v,x)\in\mathcal{V}\times\mathcal{X}$. \\
(b) If $\{\widehat{e}(v,x), \widehat{\pi}_z(v,x), \widehat{m}_z(v,x), \widehat{R}_z(v,x): z=0,1\}$ are sup-norm consistent upon $(v,x)\in\mathcal{V}\times\mathcal{X}$, and converge at rates faster than $o_p(n^{-1/4})$, 
then $n^{1/2}(\widehat\Delta_{\C}-\widehat\Delta_{\C}^*-D_n)\rightarrow_{p}0$. \\ 
(c) Further in (b), if the estimators of the parameters in $R_1(v,x;\theta_1)$ and $m_0(v,x;\theta_0)$ are regular and asymptotic linear (RAL) with influence functions $\psi_1(O)$ and $\psi_0(O)$, then $\widehat\Delta_{\C}$ is regular and asymptotic linear with influence function
$\psi(O) = \phi(O) - \Delta_{\C} + \Gamma_1\psi_1(O) - \Gamma_0\psi_0(O)$ under some regularity conditions, where
\begin{align*}
\Gamma_1 &= E\left[\frac{\partial R_1(V,X;\theta_1)}{\partial\theta_1}\{m_1(V,X)-m_0(V,X)\}\right], \\ 
\Gamma_0 &= E\left[\frac{\partial m_0(V,X;\theta_0)}{\partial\theta_0}\{R_1(V,X)-R_0(V,X)\}\right].
\end{align*}
Thus, $n^{1/2}(\widehat\Delta_{\C}-\Delta_{\C})\rightarrow_{d}N(0,E\{\psi(O)^2\})$.
\end{theorem}

Part (a) of Theorem \ref{thm3} indicates the double robustness of the AIPW type estimator. In practice, the observed survivors in the aggressive treatment group come from two principal strata, so $m_1(v,x)$ is likely to be misspecified. Double robustness provides insurance to model misspecification. Part (b) shows the higher-order bias of $\widehat{\Delta}_{\C}$ from the oracle estimator $\widehat\Delta_{\C}^{*}$. Part (c) establishes the asymptotic distribution. The proof of Theorem \ref{thm3} is given in Supplementary Material C.

\subsection{Sensitivity analysis for nondifferential substitution}

To relax the nondifferential substitution, additional knowledge regarding the behaviors of $V$ in the LD and DD strata is required. For an arbitrary $v_0\in\mathcal{V}$, let
\[
\rho(v,x) = \frac{\pi_{\text{DD}}(v,x)}{\pi_{\text{LD}}(v,x)} \Big{/} \frac{\pi_{\text{DD}}(v_0,x)}{\pi_{\text{LD}}(v_0,x)},
\]
which measures the odds ratio of DD over LD in the aggressive treatment group. Nondifferential substitution (Assumption \ref{NS}) implies that $\rho(v,x)\equiv1$. Let
\[
R^*(v,x) = \frac{\frac{\pi_1(v,x)}{1-\pi_1(v,x)}\rho(v,x)h(v,x)}{E_{V|X=x}\left\{\frac{\pi_1(V,X)}{1-\pi_1(V,X)}\rho(V,X)h(V,X)\right\}}.
\]
Without nondifferential substitution, if $\rho(v,x)$ is known, then SACEC can be identified by replacing $R(v,x)$ with $R^*(v,x)$ in Equations \eqref{SACEC1} and \eqref{SACEC2}.

\begin{remark}
If weak $S$-ignorability $S(0)\independent Z\mid V,X$ holds, then
$\pi_{\text{DD}}(v,x) = 1-\pi_1(v,x)$ and $\pi_{\text{LD}}(v,x) = \pi_1(v,x)-\pi_0(v,x)$. The nondifferential substitution assumption on the substitutional variable can be relaxed, in that $V$ can contain information of both $S(0)$ and $S(1)$. Without nondifferential substitution, taking a $v_0\in\mathcal{V}$, $\rho(v,x)$ is identified as
\[
\rho(v,x) = \frac{1-\pi_{1}(v,x)}{\pi_{1}(v,x)-\pi_{0}(v,x)} \Big{/} \frac{1-\pi_{1}(v_0,x)}{\pi_{1}(v_0,x)-\pi_{0}(v_0,x)}.
\]
\end{remark}

\begin{remark}
Although the substitutional variable $V$ plays an important role in identifying the causal estimand, interestingly, invoking the substitutional variable does not help identify $\Delta_{\C}$ under explainable nonrandom survival (also known as principal ignorability; \citealp{hayden2005estimator, ding2017principal}):
$E\{Y(1) \mid G=\text{LL}, V, X\} = E\{Y(1) \mid G=\text{LD}, V, X\}.$
The estimation of SACEC can be consistent for any misspecified $\rho(v,x)$; see Supplementary Material B. So a hypothesis test can be performed,
\begin{align*}
H_0: m_1(v_1,x)-m_0(v_1,x) = m_1(v_2,x)-m_0(v_2,x) ~\mbox{ for any }  v_1,v_2\in\mathcal{V}.
\end{align*}
If the null hypothesis $H_0$ is not rejected, then the naive (survivor-case) estimator based solely on observed surivors would not suffer from severe bias.
\end{remark}

Typically, $\rho(v,x)$ is not identifiable from observed data, which motivates a sensitivity analysis to assess the influence of $\rho(v,x)$ on the estimator. Suppose $\pi_1(v,x)$ is estimated by logistic regression, $\pi_1(v,x)/\{1-\pi_1(v,x)\}=\exp(\beta_c+x\beta_x+v\beta_v)$, and $\rho(v,x)$ is chosen as $\rho(v,x)=\exp(\eta_c+x\eta_x+v\eta_v)$, then
\[
R^*(v,x) = \frac{\exp\{v(\beta_v+\eta_v)\}h(v,x)}{\int_{\mathcal{V}}\exp\{v(\beta_v+\eta_v)\}h(v,x)p(v|x)dv}.
\]
We can introduce a sensitivity parameter $\eta\in(-1,1)$ and let $\eta_v=-\widehat\beta_v\log_2(1-\eta)$. Without loss of generality, suppose $\widehat\beta_v>0$, so $\eta_v \in (-\widehat\beta_v,\infty)$. Here we do not allow $\eta_v$ to move lower than $-\widehat\beta_v$ because we want to maintain substitution relevance.

Alternatively, we can determine the value of $\rho(v,x)$ using external interventional data; see Supplementary Material D. This setting is reasonable for post-marketing safety or effectiveness assessment based on observational data after a drug has proven beneficial for survival in some previous randomized trials.

\section{Simulation studies}\label{sec6}

In this section, we compare our proposed method with the naive survivor-case analysis and the method of \citet{wang2017identification} which requires $S$-ignorability. We generate covariates $X$ independently from
\[
\left(\begin{array}{c} X_1 \\ X_2 \end{array}\right) \sim N\left(\left(\begin{array}{c} 1 \\ 1 \end{array}\right), ~ \left(\begin{array}{cc} 1^2 & ~0.5 \\ 0.5 & ~1^2  \end{array}\right)\right), \quad X_3 \sim U(0,2),
\]
where $N(\mu,\sigma^2)$ denotes a normal distribution with mean $\mu$ and variance $\sigma^2$, and $U(l,u)$ denotes a uniform distribution on $(l,u)$. Write $X=(1,X_1,X_2,X_3)$. The substitutional variable is generated from $V \sim N(X\zeta, 2^2)$. The individualized treatment is assigned according to $\pr(Z=1\mid V,X) = \expit\{(X,V)\alpha\}$, where $\expit(x)=\exp(x)/\{1+\exp(x)\}$.

Since there may be unmeasured confounding between $Z$ and $S$, we generate the potential survivals separately in the aggressive and conservative treatment groups:
\begin{align*}
&\pr(S(0)=1\mid V, X, Z=z) = \expit\{(X,V)\beta_z\}, \\
&\pr(S(1)=1\mid V, X, Z=z, S(0)=0) = \expit\{(X,V)\gamma_z\}.
\end{align*}
The last element of $\gamma_z$ should be 0 under nondifferential substitution. Obviously, $S(1)=1$ if $S(0)=1$ according to monotonicity. The observed survival $S=ZS(1)+(1-Z)S(0)$. To account for unmeasured confounding between $Z$ and $Y$ under latent ignorability, we generate
\begin{align*}
&Y(z) \sim N\left((X,V,S(0),S(1))\delta_z, \ 1^2\right).
\end{align*}
The observed outcome $Y=ZY(1)+(1-Z)Y(0)$ if $S=1$. Set $Y=*$ (\texttt{NA}) if $S=0$.

Set $\zeta=(-1,1,0,0)^{\top}$, $\alpha=(0,-1,0,1,1)^{\top}$, $\beta_0=(-2,2,2,2,4)^{\top}$ and $\gamma_0=(-1,-1,1,-1,0)^{\top}$. Now consider four settings under which Assumptions 1--6 hold.
\begin{enumerate}
\setlength{\itemsep}{0ex}
\item Setting 1: Constant treatment effects, with $\beta_1=(2,-2,-2,-2,4)^{\top}$, $\gamma_1=(-3,1,1,1,0)^{\top}$, $\delta_0=(0,1,2,2,-2,0,-2)^{\top}$ and $\delta_1=(-1,1,2,2,-2,3,1)^{\top}$.
\item Setting 2: Ignorability (stratified randomized experiment) and exclusion restriction, with $\beta_1=\beta_0$, $\gamma_1=\gamma_0$, $\delta_0=(0,1,2,2,0,0,-2)^{\top}$ and $\delta_1=(-1,1,2,2,0,3,1)^{\top}$.
\item Setting 3: Explainable nonrandom survival and exclusion restriction, with $\beta_1=(2,-2,-2,-2,4)^{\top}$, $\gamma_1=(-3,1,1,1,0)^{\top}$, $\delta_0=(0,1,2,2,0,0,0)^{\top}$ and $\delta_1=(-1,1,2,2,0,0,1)^{\top}$.
\item Setting 4: Heterogeneous treatment effects with exclusion restriction, with $\beta_1=(2,-2,-2,-2,4)^{\top}$, $\gamma_1=(-3,1,1,1,0)^{\top}$, $\delta_0=(0,1,2,2,0,0,-2)^{\top}$ and $\delta_1=(-1,1,2,4,0,3,1)^{\top}$.
\end{enumerate}

Under ignorability (Setting 2), there is no unmeasured confounding between $Z$ and $S$. Under explainable nonrandom survival (Setting 3), there is no unmeausred confounding between $S$ and $Y$. In Settings 1 and 4, unmeasured confounding affects treatment, survival and outcome simultaneously. For each setting, we let the sample size $n\in\{200, 500, 2000\}$ and generate data 1000 times. Next, we estimate the SACEC using the naive survivor-case analysis, \citet{wang2017identification}'s method and our proposed method (regression estimator and AIPW type estimator), and then calculate the average bias, median bias and root mean squared error. The results are presented in Table \ref{tab1}.

\begin{table}
\footnotesize
\tabcolsep=0.12cm
\caption{Simulation results for estimating the SACEC}\label{tab1}
\begin{tabular}{cccccccccccccc}
\toprule
\multirow{2}{*}{Setting} & \multirow{2}{*}{$n$} & \multicolumn{4}{c}{Average bias} & \multicolumn{4}{c}{Median bias} & \multicolumn{4}{c}{Root mean squared error} \\ 
\cmidrule(lr){3-6} \cmidrule(lr){7-10} \cmidrule(lr){11-14} 
 & & SC & WZR & AIPW & REG & SC & WZR & AIPW & REG & SC & WZR & AIPW & REG \\ 
\midrule
1& 200 & -4.536 & -3.412 & -0.052 & 0.227 & -4.543 & -3.396 & -0.097 & 0.191 & 4.587 & 3.457 & 0.891 & 0.937 \\
& 500 & -4.520 & -3.374 & -0.061 & 0.376 & -4.518 & -3.368 & -0.099 & 0.342 & 4.539 & 3.391 & 0.587 & 0.742 \\
& 2000 & -4.527 & -3.388 & -0.048 & 0.617 & -4.520 & -3.387 & -0.068 & 0.595 & 4.532 & 3.392 & 0.332& 0.724 \\
2& 200 & -0.782 & 0.053 & -0.061 & 0.099 & -0.773 & 0.055 & -0.062 & 0.097 & 0.941 & 0.301 & 1.212 & 1.231 \\
& 500 & -0.773 & 0.089 & -0.022 & 0.184 & -0.778 & 0.096 & -0.069 & 0.176 & 0.836 & 0.213 & 0.746 & 0.762 \\
& 2000 & -0.783 & 0.095 & -0.014 & 0.264 & -0.784 & 0.098 & -0.043 & 0.249 & 0.800 & 0.134 & 0.526 & 0.579 \\
3& 200 & -0.722 & 0.165 & -0.010 & -0.004 & -0.702 & 0.176 & -0.012 & -0.005 & 0.898 & 0.377 & 0.615 & 0.583 \\
& 500 & -0.714 & 0.211 & 0.016 & 0.011 & -0.722 & 0.218 & 0.009 & 0.016 & 0.782 & 0.300 & 0.416 & 0.392 \\
& 2000 & -0.730 & 0.203 & -0.003 & -0.004 & -0.725 & 0.203 & -0.010 & -0.013 & 0.750 & 0.230 & 0.234 & 0.205 \\
4& 200 & -3.047 & -1.991 & -0.039 & 0.122 & -3.062 & 1.984 & -0.011 & 0.133 & 3.158 & 2.087 & 1.557 & 1.580 \\
& 500 & -3.019 & -1.944 & -0.007 & 0.199 & -3.010 & -1.922 & -0.054 & 0.180 & 3.061 & 1.980 & 0.920 & 0.942 \\
& 2000 & -3.037 & -1.955 & 0.007 & 0.285 & -3.030 & -1.958 & -0.014 & 0.283 & 3.048 & 1.964 & 0.655 & 0.712 \\
\bottomrule
\end{tabular}
SC: survivor case analysis; WZR: \citet{wang2017identification}, REG: proposed regression estimator; AIPW: proposed AIPW type 
\end{table}

The proposed AIPW type estimator is asymptotically unbiased in all these four settings. The regression estimator is biased in Settings 1, 2 and 4 due to misspecification of the outcome regression model $m_1(v,x)$, but is asymptotically unbiased in Setting 3 where $m_1(v,x)$ is correctly specified. \citet{wang2017identification}'s method is likely to be asymptotically unbiased, but prone to model misspecification, in Setting 2 where ignorability holds. The proposed AIPW type estimator has large variance in Setting 2 because it does not utilize the information of inorability. Both \citet{wang2017identification}'s method and the proposed estimators are asymptiotically unbiased in Setting 3 where explainable nonrandom survival holds, in line with Remark 3. Survivor-case analysis yields a small variance but large bias because it imposes stronger restrictions on model specifications (where the causal relationship between $G$ and $Y$ is eliminated).

Next, we conduct a sensitivity analysis. Let $\beta_1=(2,-2,-2,-2,4)^{\top}$, $\delta_0=(0,1,2,2,0,0,-2)^{\top}$, $\delta_1=(-1,1,2,2,0,3,1)^{\top}$, and $\gamma_1=(-3,1,1,1,\gamma)^{\top}$ with $\gamma\in\{-4,0,4\}$. Vary the sensitivity parameter $\eta$ and draw the average bias of 1000 repeated simulations when the sample size is 500. When $\gamma=0$, i.e., nondifferential substitution holds, the AIPW type estimator is asymptotically unbiased, with very small bias around $\eta=0$. When $\gamma=-4$, i.e., $V$ has opposite influence on $S(0)$ and $S(1)$, the AIPW type estimator is biased, and varying the sensitivity parameter $\eta$ leads to different degrees of bias. When $\gamma=4$, i.e., the influences of $V$ on $S(0)$ and $S(1)$ are in the same direction, the bias is very small varying $\eta$ around 0. As $\eta\to1$, the bias of the AIPW type estimator gets small, because $R^*(v,x)$ imposes most weights on those units with extreme values of $V$, which are highly predictive of principal strata. In contrast, \citet{wang2017identification}'s mathod and survivor-case analysis have constant but large bias. Due to model misspecification, the regression estimator is biased.

\begin{figure}
\centering
\includegraphics[width=0.32\textwidth]{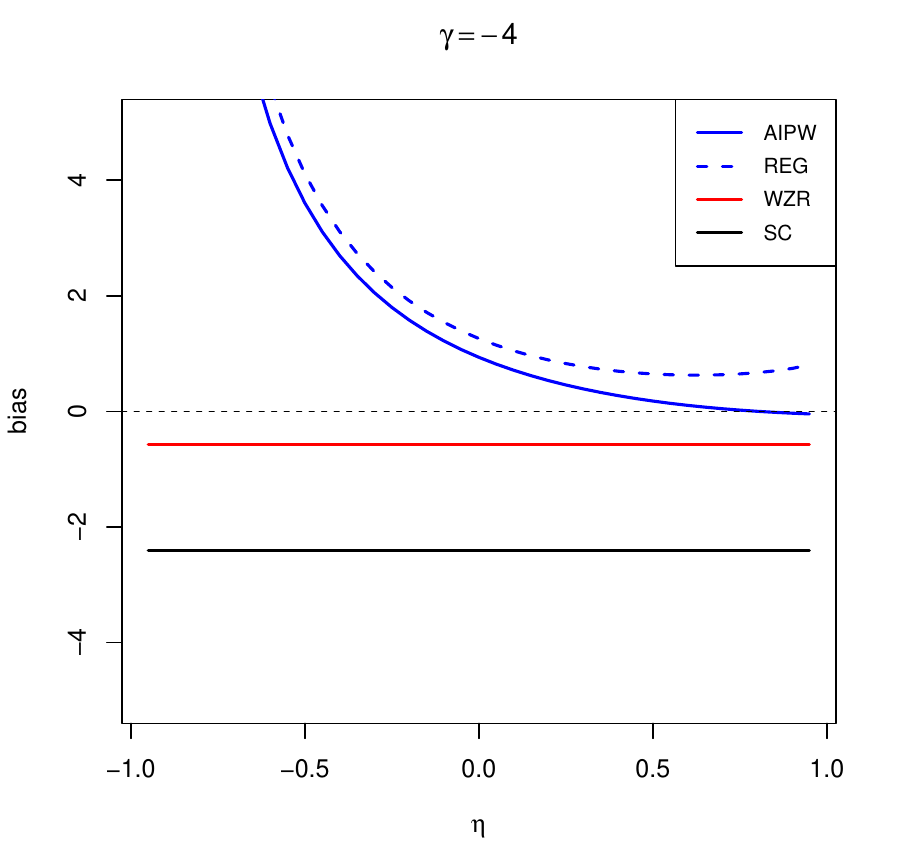}
\includegraphics[width=0.32\textwidth]{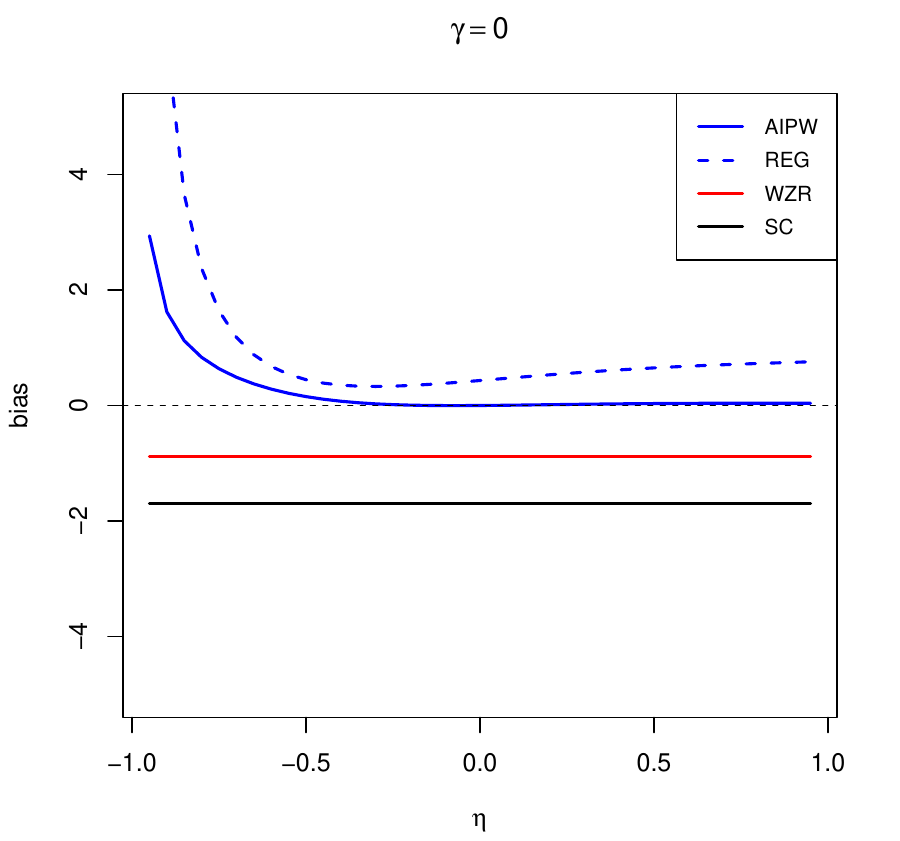}
\includegraphics[width=0.32\textwidth]{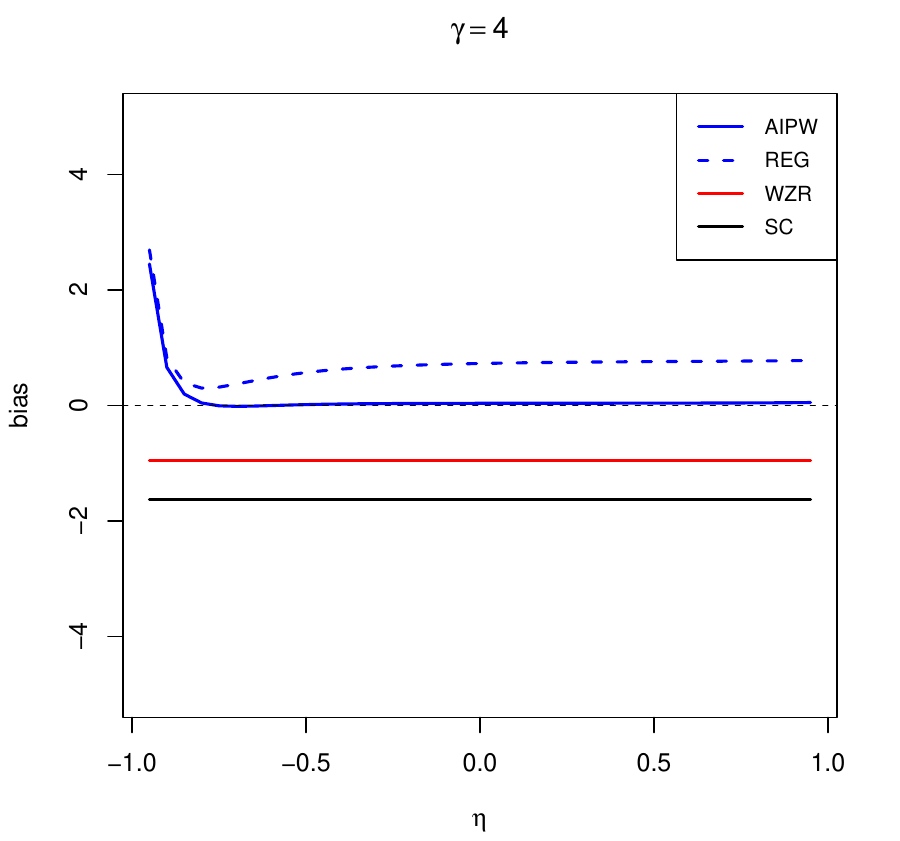}
\caption{Sensitivity analysis for nondifferential substitution. AIPW: proposed AIPW type estimator; REG: proposed regression estimatior; WZR: \citet{wang2017identification}; SC: survivor-case.}
\label{fig1}
\end{figure}

In summary, the proposed AIPW type estimator is asymptotically unbiased when Assumptions 1--6 hold and is robust to model misspecification. Even if nondifferential substitution is violated, sensitivity analysis can help assess the influence of such violation. More simulation and a sensitivity analysis for monotonicity are provided in Supplementary Material E.

\section{Application to allogeneic stem cell transplantation data}\label{sec7}

We first explain that Assumptions 1--3 hold. According to association analyses \citep{chang2020haploidentical}, three binary baseline covariates, namely minimal residual disease (MRD) presence ($X_1=1$ for positive and 0 for negative), disease status ($X_2=1$ for CR1 and 0 for CR$>$1), and diagnosis ($X_3=1$ for T-ALL and 0 for B-ALL), were found to be related with relapse. Another post-treatment biomarker, chronic graft versus host disease (GVHD), was also shown to be associated with relapse. To avoid philosophical etiology of controlling post-treatment variables, we use potential NRMs as a reflection of GVHD since GVHD is a source of NRM. Thus, it is unnatural to assume explainable nonrandom survival. Now that all risk factors of relapse are collected in $X=(X_1,X_2,X_3)$ and $(S(1),S(0))$, we expect latent ignorability to hold. Monotonicity holds because MSDT has fewer mismatched HLA loci than haplo-SCT, leading to lower NRM. Positivity holds because survival and mortality exist in both groups.

We choose age as the substitutional variable $V$. Association studies have indicated that older people have a higher probability of NRM because older people are more vulnerable to infection after surgery. However, age is not considered as a risk factor of relapse because the occurrence of relapse is more a result of biological progression. \citet{hangai2019allogeneic} found that age has a significant effect on NRM but an insignificant effect on relapse by categorizing transplantation receivers according to age. Even if age could influence relapse, it is deemed that age should have similar effects on $Y(1)$ and $Y(0)$ in all survivors because there is no biological mechanism indicating pleiotropy for age between different transplant modalities. The same result was confirmed using our dataset; see Supplementary Material F.

Table \ref{tab3} displays the estimated risk difference of transplantation types on leukemia relapse (SACEC). All methods give positive point estimates, which is a sign that haplo-SCT has a lower leukemia relapse rate than MSDT. This result is consistent with clinical experience \citep{chang2017haploidentical, chang2020haploidentical}. Although haplo-SCT leads to heavier GVHD and hence a higher probability of NRM, such GVHD inhibits progression of leukemia cells. The standard error of the AIPW type estimator is large, and the confidence interval of SACEC covers zero, probably because the substitutional variable is weak. Nevertheless, non-inferiority of haplo-SCT compared with MSDT in terms of leukemia relapse is concluded. The survivor-case analysis yields a smaller standard error because it does not consider the dependence of potential relapses on potential NRMs. \citet{wang2017identification}'s estimator may be biased due to violation of ignorability.

\begin{table}
\centering
\caption{Estimated risk difference (SACEC), with standard error (s.e.) and 95\% bootstrap confidence interval} \label{tab3}
\begin{tabular}{lcc}
\toprule
Method & SACEC estimate (s.e.) & 95\% Confidence interval \\
\midrule
Survivor-case analysis & 0.0703 (0.0298) & (0.0118, 0.1288) \\
\citet{wang2017identification} & 0.0999 (0.3020) & (-0.2022, 0.4020)  \\
Regression & 0.1522 (0.0995) & (-0.0428, 0.3472) \\
AIPW type & 0.1292 (0.1291) & (-0.1240, 0.3824) \\
\bottomrule
\end{tabular}
\end{table}

\begin{figure}
\centering
\includegraphics[width=0.6\textwidth]{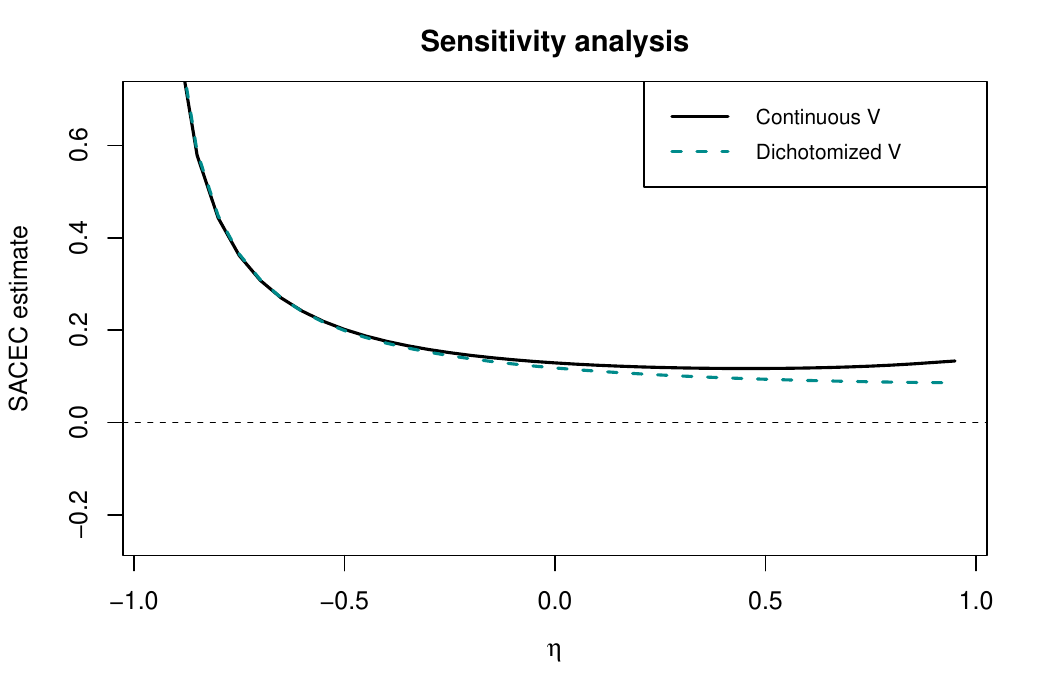}
\caption{Sensitivity analysis of the AIPW type estimator on the stem cell transplantation data.}
\label{fig2}
\end{figure}

To perform a sensitivity analysis, Figure \ref{fig2} displays the curve of estimated SACEC varying the sensitivity parameter $\eta\in(-1,1)$. We also tried a dichotomized $V$ cut at its mean to assess the robustness of the SACEC estimate to $V$. The point estimates of SACEC are always positive, indicating that the substantive conclusion is insensitive to moderate deviation from the nondifferential substitution assumption. In a nutshell, haplo-SCT is a non-inferior alternative to MSDT in terms of leukemia relapse. With the maturing intervention on survival, haplo-SCT could be a promising choice in the future.

\section{Discussion}\label{sec8}

In this study, we examined the identification and estimation of causal effects when outcomes are truncated by death, where both (a) the treatment assignment and survival process, and (b) the survival process and outcome process are confounded. It is not straightforward to extend the identification results for the SACE directly from randomized experiments to observational data because the proportion of always-survivors is not always identifiable. We emphasize that the violation of explainable nonrandom survival (principal ignorability) is the key obstacle to identifying the causal effect. A substitutional variable for $S(0)$, i.e., a proxy of always-survivorship, is required to recover the information regarding the principal strata. The proposed method significantly improves the utility of real-world data and provides convenience for post-marketing drug safety and effectiveness assessments.

The introduction of a substitutional variable is similar to proximal causal inference using negative control variables \citep{tchetgen2020introduction, cui2020semiparametric} or auxiliary variables \citep{miao2022identifying}, but there are some important differences. First, the target estimand in the truncation-by-death setting is the local effect (i.e., the causal effect in always-survivors) rather than the overall effect. Second, we only use one variable related with the principal strata to recover the information in the principal strata. With only a single substitutional variable, the choice can be more flexible. Third, the substitutional variable is for the principal strata rather than unmeasured confounders. Finding a risk factor of $S(0)$ is easier than finding proxies of unmeasured confounders which must be rich enough to recover all the information in unmeasured confounders. If there are several eligible substitutional variables, all these variables can be involved in the regressions.

The proposed method to identify the SACEC still bears some limitations becasue it relies on untestable assumptions. Latent ignorability implies that the confounding effect between the treatment assignment and outcome process can be adjusted by principal strata and observed covariates. Latent ignorability is likely to hold if as many predictors of the treatment and outcome as possible are controlled. The nondifferential substitution assumption may be strong in some scenarios. Sensitivity analysis should be conducted when there are no external information to determine the survivorship odds ratio $\rho(v,x)$. The assumptions we adopt are the minimum requirements to identify causal effects with a single substitutional variable. Although it is possible to achieve identification of the constitution of principal strata without nondifferential substitution by double proxies for the unmeasured confounder $U$, other assumptions on the proxies like exclusion restriction or confounding bridge should be made instead \citep{luo2022identification}.

\section*{Acknowledgements}

We thank Shanshan Luo and Thomas S. Richardson for discussion. We also thank Leqing Cao for cleaning the data. The methods were implemented using R version 4.2.1. \emph{Funding information:} National Key Research and Development Program of China, Grant No.~2021YFF0901400; National Natural Science Foundation of China, Grant No.~12026606; Novo Nordisk A/S.

\section*{Supplementary Material}

The Supplememtary Material, which includes causal graphs, proofs of theorems, additional simulation and data analysis results, is available online. Data and R codes are available at https://github.com/naiiife/TruncDeath.

\bibliographystyle{apalike}
\bibliography{paper-ref}



\section*{Supplementary material}

# Supplementary Material for "Causal Inference with Truncation-by-Death and Unmeasured Confounding"

**Abstract**

This supplementary material consists of six parts: (A) estimands and examples; (B) proofs of identification theorems; (C) proofs of large sample properties of estimators; (D) comments on how to relax the nondifferential substitution by external data; (E) more simulation results; (F) assumption checks in real data.

## A  Estimands and examples

### A.1  Estimands

Stratified by the joint values of potential survivals, there are four principal strata, as listed in Table S1. Since the potential survivals are unaffected by actual treatments, the principal strata act as baseline covariates characterizing different features of individuals (Jiang and Ding, 2021). We should restrict our analysis to the LL stratum because both potential outcomes are well defined only in the LL stratum. When monotonicity holds, the DL stratum is excluded. To facilitate understanding, we recode the treatment with higher survival as the aggressive treatment ($Z = 1$), and the one with lower survival as the conservative treatment ($Z = 0$).

Table S1: Principal strata (✓ means well defined and ∗ means undefined)

| Stratum | $S(1)$ | $S(0)$ | $Y(1)$ | $Y(0)$ | Description | |
|---|---|---|---|---|---|---|
| LL | 1 | 1 | ✓ | ✓ | Always-survivors | |
| LD | 1 | 0 | ✓ | ∗ | Protected | |
| DL | 0 | 1 | ∗ | ✓ | Harmed | (Excluded by monotonicity) |
| DD | 0 | 0 | ∗ | ∗ | Doomed | |

To be noted, although the aggressive treatment leads to higher survival than the conservative treatment, the relation on the levels of primary outcomes does not necessarily satisfy monotonicity. The aggressive treatment may result in worse quality of life than the conservative treatment. In many practical situations, practitioners want to know whether the conservative treatment is sufficient. If the conservative treatment can lead to better quality of life, it could be more worthy to give patients the conservative treatment as long as they can survive.

Define the conditional survivor average causal effect (CSACE)

$$\Delta(v, x) = E\{Y(1) - Y(0) \mid G = \text{LL}, V = v, X = x\}.$$

For a unit with covariates $(v, x)$, this quantity measures how large the treatment effect would be if the unit could survive regardless of the treatment. Consider the conditional survivor average causal effect on the aggressive treatment (CSACET) and the conditional survivor average causal effect on the conservative treatment (CSACEC) as follows,

$$\Delta_{\text{T}}(v, x) = E\{Y(1) - Y(0) \mid G = \text{LL}, Z = 1, V = v, X = x\},$$

1

$$\Delta_{\mathrm{C}}(v, x) = E\{Y(1) - Y(0) \mid G = \mathrm{LL}, Z = 0, V = v, X = x\}.$$

To average the covariates out, define the survivor average causal effect (SACE)

$$\Delta = E\{Y(1) - Y(0) \mid G = \mathrm{LL}\}.$$

Similarly, we can define the survivor average causal effect on the aggressive treatment (SACET) and the survivor average causal effect on the conservative treatment (SACEC) as follows,

$$\Delta_{\mathrm{T}} = E\{Y(1) - Y(0) \mid G = \mathrm{LL}, Z = 1\},$$
$$\Delta_{\mathrm{C}} = E\{Y(1) - Y(0) \mid G = \mathrm{LL}, Z = 0\}.$$

Under monotonicity, the observed survivors in the conservative treatment group must come from the LL stratum. Therefore, $\Delta_{\mathrm{C}}$ is an estimand defined on an individually observable subpopulation. Since the proportion of always-survivors is not identifiable in the aggressive treatment group, $\Delta_{\mathrm{T}}$ and $\Delta$ are not identifiable.

Next, introduce the propensity score, survival scores and outcome regression models,

$$e(v, x) = \mathrm{pr}(Z = 1 \mid V = v, X = x),$$
$$\pi_0(v, x) = \mathrm{pr}(S = 1 \mid Z = 0, V = v, X = x),$$
$$\pi_1(v, x) = \mathrm{pr}(S = 1 \mid Z = 1, V = v, X = x),$$
$$m_0(v, x) = E(Y \mid Z = 0, S = 1, V = v, X = x),$$
$$m_1(v, x) = E(Y \mid Z = 1, S = 1, V = v, X = x).$$

These quantities can be estimated from data by generalized linear models or nonparametric models.

## A.2 Examples

In the following we give some examples to explain the meanings of assumptions.

*Example 1.* Suppose a traditional therapy (such as chemotherapy) has been widely adopted to cure a disease, but this therapy may cause adverse effects (readmission). Now a novel therapy (such as medication) is invented, which can greatly lower the safety risk, so monotonicity holds. In a cohort study, patients decided proper treatments to take based on doctor's suggestion. Let $Z$ be the treatment a patient received, $S$ be the occurence of readmission, and $Y$ be the outcome of interest, such as quality of life. We want to measure the treatment effect on quality of life at home. The substitutional variable $V$ can be a baseline risk factor of the adverse effect for the chemotherapy. By definition, $V$ satisfies substitution relevance and nondifferential substitution since $V$ is irrelevant to medication outcomes. Suppose the risk factor $V$ would not affect the quality of life if this patient can live normally outside the hospital, so exclusion restriction (non-interaction) is also satisfied.

*Example 2.* Suppose a targeted durg is used to cure cancer. This targeted drug is effective only if the patient carry certain gene. Now a generic drug targeting on a wider range of but unconfirmed population is invented. Let $Z$ indicate which drug to take, $S$ be an indicator indicating whether a patient response to the drug, and $Y$ be the outcome of interest. Practitioners want to know whether the generic drug has a better performance than the targeted drug. To make these two drugs comparative, the study should condition on the subpopulation who are responsive to both drugs. In a clinical trial, patients with PCR testing positive for the targeted gene are randomized to take the targeted drug or generic drug, while patients with PCR testing negative are all assigned to take the generic drug. Monotonicity means that the patients with PCR testing positive are also eligible for the generic drug, ensured by the wider target of the generic drug. The substitutional variable $V$ can be the PCR test result. It satisfies substitution

relevance and nondifferential substitution because the PCR test cannot predict the response to the generic drug if a patient does not respond to the targeted drug. It also satisfies non-interaction because PCR test results cannot directly influence outcomes: it is $G = (S(1), S(0))$ that may influence outcomes.

*Example 3.* Suppose the government puts forward a training program to promote employment for job finders. Here $Z$ is whether a person joins in the program, $S$ is an indicator of employment, and $Y$ is the earning if employed. The government wants to know the effect of this program on earnings, so the study should restrict to the always-employed subpopulation. A person who tends to be unemployed would be more willing to join in the program. Monotonicity means that joining in the program would make job finding easier. The substitutional variable $V$ can be self-reported confidence of successfully finding a job before joining in the program, so $V$ satisfies substitution relevance and nondifferential substitution. Such a variable $V$ can be collected by questionnaire when signing up for the program. Non-interaction is also reasonable because once he/she is employed, the pre-treatment confidence for being employed measured before joining in the training program would not make a difference on earnings whether he/she joined in the training program or not.

## A.3 Causal graphs illustration

To demonstrate the relation of variables, we use the language of causal graphs for an intuitive illustration.

The principal strata variable $G$ can be expressed as $(S(1), S(0))$, so $G$ and $Z$ together determines the observed $S$. Figure S1 shows three situations with unmeasured confounding. The substitutional variable $V$ is omitted on the graphs. The subfigure (a) is a case where the survival process and the outcome process are confounded by principal strata, which was considered by Wang et al. (2017). A substitutional variable $V$ related with $G$ is needed to identify the conditional survivor average causal effect. The subfigure (b) is a case where the treatment assignment and the survival process are confounded by principal strata, which was considered by Egleston et al. (2009). Following the illustration in Kédagni (2021), it can be considered that there is an unmeasured confounder $U_G$ having direct effects on $Z$ and $G$. The absence of $G \to Y$ edge implies explainable nonrandom survival, in that the outcomes in survivors are unaffected by principal strata. Conditioning on $S$, the effect of the unmeasured $U_G$ or $G$ on $Y$ is blocked by $S$. Therefore, the conditional survivor average causal effect is directly identified based on observed survivors. The subfigure (c) is a case where both the treatment–survival processes and the survival–outcome processes are confounded by principal strata. Similar with (b), the unmeasured confounder $U_G$ can be absorbed into $G$ since both $U_G$ and $G$ are unobservable. A substitutional variable related with $G$ (which should also satisfy other assumptions) is needed to identify the conditional survivor average causal effect.

When the treatment–survival–outcome processes are confounded by principal strata and covariates (latent ignorability), as in subfigure (c) of Figure S1, the survivor-case analysis is biased without the explainable nonrandom survival. Conditioning on the observed survivors $S = 1$ and covariates $X$, there emerges a spurious effect between $Z = z$ and $G$ (for analysis on causal graphs, see Verma and Pearl, 1988, Glymour and Greenland, 2008 or Elwert, 2013). If the explainable nonrandom survival assumption does not hold, the spurious effect of $Z = z$ can be delivered through the open path from $G$ to $Y$. As a result, the survivor-case analysis leads to a biased estimator of the true treatment effect. To block all the backdoor paths between $Z$ and $Y$, the principal strata $G$ and covariates $(V, X)$ must be controlled. Thus, $\Delta(v, x) = \Delta_\text{T}(v, x) = \Delta_\text{C}(v, x)$. By introducing a substitutional variable, $\Delta(v, x)$ can be identified (where $V$ acts like an instrumental variable for $G$). However, it is hard to identify the overall survivor average causal effect. The edge $X \to G$ is unpredictable since $G$ is unobserved, unless conditioning on $Z = 0$ and $S = 1$ which implies $G = \text{LL}$. Since $G = \text{LL}$ can be recovered in the conservative treatment group, $\Delta_\text{C}$ is identifiable.

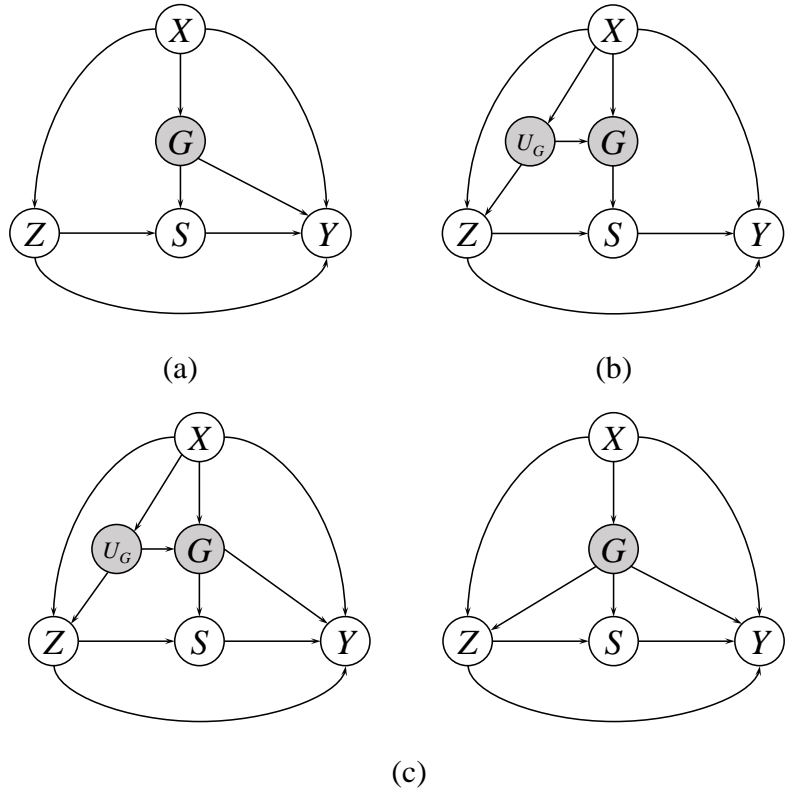

Figure S1: Causal graphs, when (a) the survival–outcome processes are confounded by principal strata, (b) the treatment–survival processes are confounded by principal strata, and (c) the treatment–survival–outcome processes are confounded by principal strata. The substitutional variable $V$ is omitted.

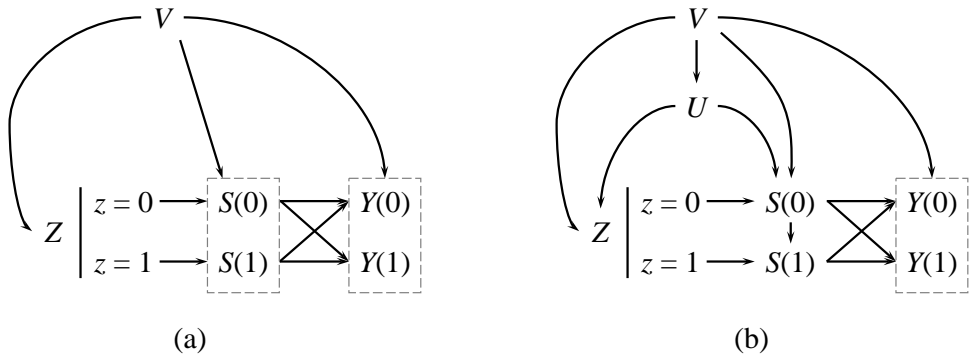

Figure S2: Causal structural models, with (a) ignorability or (b) latent ignorability. Baseline covariates $X$, which can have edges to $V$, $Z$, $U$, $S(0)$, $S(1)$, $Y(0)$ and $Y(1)$, are omitted.

4

To explain latent ignorability, Figure S2 shows the causal structural models, displaying the potential values on the causal graphs. Baseline covariates $X$, which can have edges to $V$, $Z$, $U$, $S(0)$, $S(1)$, $Y(0)$ and $Y(1)$, are omitted for brevity. Suppose there is an unmeasured confounder $U$. In randomized experiments or observational studies with ignorability

$$Z \perp\!\!\!\perp (S(0), S(1), Y(0), Y(1)) \mid V, X,$$

the influence of the unmeasured confounding on the treatment assignment $Z$ is blocked by controlling observed covariates $(V, X)$, as shown in subfigure (a). Given the principal strata $G = (S(1), S(0))$, we assume that the risk factors of $Y$ are all controlled in $(V, X)$. To be noted, ignorability does not exclude the cross-world reliance of $Y(z)$ on $S(1-z)$. Ignorability implies latent ignorability

$$Z \perp\!\!\!\perp (Y(0), Y(1)) \mid S(0), S(1), V, X.$$

Specially, we have $Z \perp\!\!\!\perp Y(z) \mid G = \text{LL}, V, X$ if we let $S(0) = S(1) = 1$, for $z = 0, 1$. We consider the expectation of $Y(z)$ as a function of $(S(1), S(0))$ and $(V, X)$, probably with exogenous noises. Latent ignorability is weaker than ignorability in that latent ignorability allows unmeasured common causes for $Z$ and $(S(1), S(0))$, as shown in subfigures (b) and (c). However, given $G = \text{LL}$ and covariates $(V, X)$, there would be no unmeasured common causes for $Z$ and $Y(z)$.

The substitutional variable $V$ should be correlated with $S(0)$. Nondifferential substitution means that $V$ is a proxy of $S(0)$, whereas the influence of $V$ on $S(1)$ is nondifferential (analogous to the nondifferential measurement error assumption; Carroll and Stefanski, 1990; Guo et al., 2012). Given $S(0) = 0$ and covariates $X$, then $V$ would be independent of $S(1)$. For example, $U$ is the parent of $S(0)$ and $V$ is the parent of $U$. Or, $V$ is a risk factor of $S(0)$, the natural level of survival, and the treatment gives a lift on survival upon $S(0)$, resulting in $S(1)$. Moreover, as in the non-interaction assumption, $V$ should not modify the treatment effect on the primary outcome in any principal stratum.

## B  Proofs of identification theorems

### B.1  To identify the conditional survivor average causal effect (CSACE)

We first show that $\Delta(v, x) = \Delta(x)$, a function irrelevant to $v$. By non-interaction (Assumption 6), we can find functions $\mu_z(x)$, $z = 0, 1, 2$, so that

$$\mu_1(v, x) := E\{Y(1) \mid Z = 1, G = \text{LL}, V = v, X = x\} = \mu_1(x) + f(v, x),$$
$$\mu_2(v, x) := E\{Y(1) \mid Z = 1, G = \text{LD}, V = v, X = x\} = \mu_2(x) + f(v, x),$$
$$\mu_0(v, x) := E\{Y(0) \mid Z = 0, G = \text{LL}, V = v, X = x\} = \mu_0(x) + f(v, x).$$

In the conservative treatment group $(Z = 0)$,

$$\mu_0(v, x) = E\{Y(0) \mid Z = 0, G = \text{LL}, V = v, X = x\}$$
$$= E\{Y \mid Z = 0, S = 0, V = v, X = x\} = m_0(v, x)$$

is directly identifiable from observed data. By latent ignorability (Assumption 1),

$$E\{Y(1) \mid G = \text{LL}, V = v, X = x\} = \mu_1(v, x) = \mu_1(x) + f(v, x),$$
$$E\{Y(0) \mid G = \text{LL}, V = v, X = x\} = \mu_0(v, x) = \mu_0(x) + f(v, x).$$

Therefore,

$$\Delta(v, x) = E\{Y(1) - Y(0) \mid G = \text{LL}, V = v, X = x\} = \mu_1(x) - \mu_0(x) =: \Delta(x).$$

Moreover, the conditional survivor average causal effects in both treatment groups are irrelevant to $v$ by latent ignorability, i.e., $\Delta(x) = \Delta_{\text{T}}(x) = \Delta_{\text{C}}(x)$.

**Lemma 1** *Under Assumptions 1 to 6, CSACE can be identified. Let $h(v,x)$ be an integrable function with respect to $v$ subject to*

$$\int_{\mathcal{V}} h(v,x)p(v|x)dv = 0, \quad \int_{\mathcal{V}} \frac{\pi_{LL}(v,x)}{\pi_{LD}(v,x)} h(v,x)p(v|x)dv \neq 0$$

*for every $x \in \mathcal{X}$, and $p(v|x)$ be the conditional density of $V$ given $X = x$. Then*

$$\Delta(x) = \frac{\int_{\mathcal{V}} \frac{\pi_1(v,x)}{1-\pi_1(v,x)} \{m_1(v,x) - m_0(v,x)\} h(v,x) p(v|x) dv}{\int_{\mathcal{V}} \frac{\pi_1(v,x)}{1-\pi_1(v,x)} h(v,x) p(v|x) dv} \quad (1)$$

*given that the right side is finite.*

**Proof sketch of Lemma S1.** Without loss of generality, suppose $\mathcal{V} = \{0,1\}$ Let $\pi_g(v,x) = \mathrm{pr}(G = g \mid Z = 1, V = v, X = x)$, for $g \in \{\mathrm{LL}, \mathrm{LD}, \mathrm{DD}\}$. By the Bayesian formula,

$$\mathrm{pr}(V = v \mid Z = 1, G = g, X = x) = \frac{\mathrm{pr}(V = v \mid Z = 1, X = x)\pi_g(v,x)}{\sum_{u=0}^{1} \mathrm{pr}(V = u \mid Z = 1, X = x)\pi_g(u,x)}.$$

By nondifferential substitution (Assumption 5),

$$\mathrm{pr}(V = v \mid Z = 1, G = \mathrm{LD}, X = x) = \mathrm{pr}(V = v \mid Z = 1, G = \mathrm{DD}, X = x).$$

This implies

$$\frac{\pi_{\mathrm{DD}}(1,x)}{\pi_{\mathrm{LD}}(1,x)} = \frac{\pi_{\mathrm{DD}}(0,x)}{\pi_{\mathrm{LD}}(0,x)} =: r(x)$$

for some unknown value $r(x)$. For $v = 0, 1$, we have

$$E(YS \mid Z = 1, V = v, X = x) = \pi_{\mathrm{LL}}(v,x)\{\mu_1(x) + f(v,x)\} + \pi_{\mathrm{LD}}(v,x)\{\mu_2(x) + f(v,x)\},$$
$$E(S \mid Z = 1, V = v, X = x) = \pi_{\mathrm{LL}}(v,x) + \pi_{\mathrm{LD}}(v,x).$$

Recall that

$$E(Y \mid Z = 0, S = 1, V = v, X = x) = \mu_0(x) + f(v,x),$$

so we can solve $\mu_1(x) - \mu_0(x) = \{\mu_1(x) + f(v,x)\} - \{\mu_0(x) + f(v,x)\}$. $\square$

**Proof of Lemma S1.** By nondifferential substitution,

$$r(x) = \frac{\pi_{\mathrm{DD}}(v,x)}{\pi_{\mathrm{LD}}(v,x)}$$

does not depend on $v$. By monotonicity,

$$\pi_{\mathrm{DD}}(v,x) = 1 - \pi_{\mathrm{LL}}(v,x) - \pi_{\mathrm{LD}}(v,x) = 1 - \pi_1(v,x).$$

Since

$$\pi_1(v,x)m_1(v,x)$$
$$= \mathrm{pr}(S = 1 \mid Z = 1, V = v, X = x)E(Y \mid Z = 1, S = 1, V = v, X = x)$$
$$= E(YS \mid Z = 1, V = v, X = x)$$
$$= \pi_{\mathrm{LL}}(v,x)E(YS \mid Z = 1, G = \mathrm{LL}, V = v, X = x)$$
$$\quad + \pi_{\mathrm{LD}}(v,x)E(YS \mid Z = 1, G = \mathrm{LD}, V = v, X = x)$$
$$\quad + \pi_{\mathrm{DD}}(v,x)E(YS \mid Z = 1, G = \mathrm{DD}, V = v, X = x)$$
$$= \pi_{\mathrm{LL}}(v,x)E\{Y(1) \mid Z = 1, G = \mathrm{LL}, V = v, X = x\}$$
$$\quad + \pi_{\mathrm{LD}}(v,x)E\{Y(1) \mid Z = 1, G = \mathrm{LD}, V = v, X = x\}$$

$$= \pi_{\mathrm{LL}}(v,x)\{\mu_1(x) + f(v,x)\} + \pi_{\mathrm{LD}}(v,x)\{\mu_2(x) + f(v,x)\},$$

it follows that

$$m_1(v,x) = \frac{\pi_{\mathrm{LL}}(v,x)}{\pi_1(v,x)}\{\mu_1(x) + f(v,x)\} + \frac{\pi_{\mathrm{LD}}(v,x)}{\pi_1(v,x)}\{\mu_2(x) + f(v,x)\}.$$

Note that $\pi_{\mathrm{LL}}(v,x) + \pi_{\mathrm{LD}}(v,x) = \pi_1(v,x)$, so

$$m_0(v,x) = \frac{\pi_{\mathrm{LL}}(v,x)}{\pi_1(v,x)}\{\mu_0(x) + f(v,x)\} + \frac{\pi_{\mathrm{LD}}(v,x)}{\pi_1(v,x)}\{\mu_0(x) + f(v,x)\}.$$

Thus,

$$\int_{\mathcal{V}} \frac{\pi_1(v,x)}{1-\pi_1(v,x)}\{m_1(v,x) - m_0(v,x)\}h(v,x)p(v|x)dv$$
$$= \int_{\mathcal{V}} \frac{\pi_{\mathrm{LL}}(v,x)}{\pi_{\mathrm{DD}}(v,x)}\{\mu_2(x) - \mu_0(x)\}h(v,x)p(v|x)dv + \int_{\mathcal{V}} \frac{\pi_{\mathrm{LD}}(v,x)}{\pi_{\mathrm{DD}}(v,x)}\{\mu_2(x) - \mu_0(x)\}h(v,x)p(v|x)dv$$
$$= \{\mu_1(x) - \mu_0(x)\}r(x)^{-1}\int_{\mathcal{V}} \frac{\pi_{\mathrm{LL}}(v,x)}{\pi_{\mathrm{LD}}(v,x)}h(v,x)p(v|x)dv + \{\mu_2(x) - \mu_0(x)\}r(x)^{-1}\int_{\mathcal{V}} h(v,x)p(v|x)dv$$
$$= \{\mu_1(x) - \mu_0(x)\}r(x)^{-1}\int_{\mathcal{V}} \frac{\pi_{\mathrm{LL}}(v,x)}{\pi_{\mathrm{LD}}(v,x)}h(v,x)p(v|x)dv,$$

and

$$\int_{\mathcal{V}} \frac{\pi_1(v,x)}{1-\pi_1(v,x)}h(v,x)p(v|x)dv$$
$$= \int_{\mathcal{V}} \frac{1}{\pi_{\mathrm{DD}}(v,x)}\{\pi_{\mathrm{LL}}(v,x) + \pi_{\mathrm{LD}}(v,x)\}h(v,x)p(v|x)dv$$
$$= \int_{\mathcal{V}} \frac{\pi_{\mathrm{LL}}(v,x)}{\pi_{\mathrm{DD}}(v,x)}h(v,x)p(v|x)dv + \int_{\mathcal{V}} \frac{\pi_{\mathrm{LD}}(v,x)}{\pi_{\mathrm{DD}}(v,x)}h(v,x)p(v|x)dv$$
$$= r(x)^{-1}\int_{\mathcal{V}} \frac{\pi_{\mathrm{LL}}(v,x)}{\pi_{\mathrm{LD}}(v,x)}h(v,x)p(v|x)dv + r(x)^{-1}\int_{\mathcal{V}} h(v,x)p(v|x)dv$$
$$= r(x)^{-1}\int_{\mathcal{V}} \frac{\pi_{\mathrm{LL}}(v,x)}{\pi_{\mathrm{LD}}(v,x)}h(v,x)p(v|x)dv.$$

Therefore,

$$\Delta(x) = \mu_1(x) - \mu_0(x) = \frac{\int_{\mathcal{V}} \frac{\pi_1(v,x)}{1-\pi_1(v,x)}\{m_1(v,x) - m_0(v,x)\}h(v,x)p(v|x)dv}{\int_{\mathcal{V}} \frac{\pi_1(v,x)}{1-\pi_1(v,x)}h(v,x)p(v|x)dv}.$$

If we write

$$R(v,x) = \frac{\frac{\pi_1(v,x)}{1-\pi_1(v,x)}h(v,x)}{\int_{\mathcal{V}} \frac{\pi_1(v,x)}{1-\pi_1(v,x)}h(v,x)p(v|x)dv},$$

then

$$\Delta(x) = E_{V|X=x}\left[R(V,X)\{m_1(V,X) - m_0(V,X)\}\right].$$

The conditional survivor average causal effect (CSACE) is identified. □

More generally, it can be shown that

$$\Delta(x) = E_{V|X=x}\left[R(V,X)\{m_1(V,X) - f(V,X)\}\right] - \{m_0(v,x) - f(v,x)\}.$$

If the nondifferential substitution assumption is relaxed with

$$\rho(v,x) = \frac{\pi_{\mathrm{DD}}(v,x)}{\pi_{\mathrm{LD}}(v,x)} \bigg/ \frac{\pi_{\mathrm{DD}}(v_0,x)}{\pi_{\mathrm{LD}}(v_0,x)} = \frac{\pi_{\mathrm{DD}}(v,x)}{\pi_{\mathrm{LD}}(v,x)}\rho_0^{-1}(x),$$

7

it can be shown in the same way that

$$\int_{\mathcal{V}} \frac{\pi_1(v,x)}{1-\pi_1(v,x)} \rho(v,x) h(v,x) p(v|x) dv$$
$$= \rho_0^{-1}(x) \int_{\mathcal{V}} \frac{\pi_{\mathrm{LL}}(v,x)}{\pi_{\mathrm{LD}}(v,x)} h(v,x) p(v|x) dv,$$

$$\int_{\mathcal{V}} \frac{\pi_1(v,x)}{1-\pi_1(v,x)} \rho(v,x)\{m_1(v,x) - m_0(v,x)\} h(v,x) p(v|x) dv$$
$$= \{\mu_1(x) - \mu_0(x)\} \rho_0^{-1}(x) \int_{\mathcal{V}} \frac{\pi_{\mathrm{LL}}(v,x)}{\pi_{\mathrm{LD}}(v,x)} h(v,x) p(v|x) dv,$$

so

$$\Delta(x) = \mu_1(x) - \mu_0(x) = \frac{\int_{\mathcal{V}} \frac{\pi_1(v,x)}{1-\pi_1(v,x)}\{m_1(v,x) - m_0(v,x)\}\rho(v,x)h(v,x)p(v|x)dv}{\int_{\mathcal{V}} \frac{\pi_1(v,x)}{1-\pi_1(v,x)}\rho(v,x)h(v,x)p(v|x)dv}.$$

## B.2 Blessings of explainable nonrandom survival

Although the substitutional variable $V$ plays an important role to identify the target causal estimands, interestingly, invoking the substitutional variable would not help to identify $\Delta(x)$ under the explainable nonrandom suvival (principal ignorability). In fact, explainable nonrandom survival assumes that the conditional expectation of $Y(z)$ is not affected by the cross-world potential survival $S(1-z)$,

$$m_1(v,x) = E(Y \mid Z=1, S=1, V=v, X=x)$$
$$= E\{Y(1) \mid Z=1, G=\mathrm{LL}, V=v, X=x\} = \mu_1(x) + f(v,x)$$
$$= E\{Y(1) \mid Z=1, G=\mathrm{LD}, V=v, X=x\} = \mu_2(x) + f(v,x).$$

That is, $\mu_1(x) = \mu_2(x)$. As a result,

$$\frac{\int_{\mathcal{V}} \frac{\pi_1(v,x)}{1-\pi_1(v,x)}\{m_1(v,x) - m_0(v,x)\}\rho(v,x)h(v,x)p(v|x)dv}{\int_{\mathcal{V}} \frac{\pi_1(v,x)}{1-\pi_1(v,x)}\rho(v,x)h(v,x)p(v|x)dv}$$
$$= \{\mu_1(x) - \mu_0(x)\} \frac{\int_{\mathcal{V}} \frac{\pi_1(v,x)}{1-\pi_1(v,x)}\rho(v,x)h(v,x)p(v|x)dv}{\int_{\mathcal{V}} \frac{\pi_1(v,x)}{1-\pi_1(v,x)}\rho(v,x)h(v,x)p(v|x)dv}$$
$$= \mu_1(x) - \mu_0(x)$$
$$= \Delta(x)$$

for any $\rho(v,x)$.

## B.3 To identify the survivor average causal effect on the conservative treatment (SACEC)

**Proof of Theorem 1.** Since $\Delta(x) = \Delta_{\mathrm{C}}(x)$ has been identified, by the total probability formula,

$$\Delta_{\mathrm{C}} = E\{Y(1) - Y(0) \mid Z=0, G=\mathrm{LL}\}$$
$$= \int_{\mathcal{X}} \Delta(x) \mathrm{pr}(X=x \mid Z=0, G=\mathrm{LL}) dx$$
$$= \int_{\mathcal{X}} \Delta(x) \frac{\mathrm{pr}(Z=0, G=\mathrm{LL} \mid X=x)p(x)}{\mathrm{pr}(Z=0, G=\mathrm{LL})} dx$$
$$= \int_{\mathcal{X}} \Delta(x) \frac{\mathrm{pr}(Z=0, S=1 \mid X=x)p(x)}{\mathrm{pr}(Z=0, S=1)} dx,$$

where $p(x)$ is the density of $X$. Write

$$R_0(v,x) = \frac{(1-e(v,x))\pi_0(v,x)}{\text{pr}(Z=0,S=1)} = \frac{\text{pr}(Z=0,S=1 \mid V=v, X=x)}{\text{pr}(Z=0,S=1)},$$

then

$$E_{V|X=x}\{R_0(V,X)\} = \frac{\text{pr}(Z=0,S=1 \mid X=x)}{\text{pr}(Z=0,S=1)},$$

so

$$\begin{aligned}
\Delta_\text{C} &= \int_\mathcal{X} \Delta(x) E_{V|X=x}\{R_0(V,X)\} dx = E_X\left[\Delta(X) E_{V|X}\{R_0(V,X)\}\right] \\
&= E_X\left\{E_{V|X}[R(V,X)\{m_1(V,X) - m_0(V,X)\}] E_{V|X}\{R_0(V,X)\}\right\} \\
&= E_X\left\{E_{V|X}[R(V,X) E_{V|X}\{R_0(V,X)\}\{m_1(V,X) - m_0(V,X)\}]\right\} \\
&= E_X\left\{E_{V|X}[R_1(V,X)\{m_1(V,X) - m_0(V,X)\}]\right\} \\
&= E_X\left[R_1(V,X)\{m_1(V,X) - m_0(V,X)\}\right]. 
\end{aligned} \qquad (2)$$

Moreover, $\Delta_\text{C}$ can be identified by an augmented inverse probability weighting (AIPW) type formula. Note that

$$\begin{aligned}
&E\left[R_1(V,X)\left\{\frac{ZS(Y-m_1(V,X))}{e(V,X)\pi_1(V,X)} + m_1(V,X) - m_0(V,X)\right\}\right] \\
&= E\left\{E\left[R_1(V,X)\left\{\frac{ZS(Y-m_1(V,X))}{e(V,X)\pi_1(V,X)} + m_1(V,X) - m_0(V,X)\right\}\bigg| Z,S,V,X\right]\right\} \\
&= E\left\{E\left[R_1(V,X)\left\{m_1(V,X) - m_0(V,X)\right\} \bigg| Z,S,V,X\right]\right\} \\
&= E\left[R_1(V,X)\left\{m_1(V,X) - m_0(V,X)\right\}\right] = \Delta_\text{C},
\end{aligned}$$

and

$$\begin{aligned}
&E\left[R_0(V,X)\left\{\frac{(1-Z)S(Y-m_0(V,X))}{(1-e(V,X))\pi_0(V,X)}\right\}\right] \\
&= E\left\{E\left[R_0(V,X)\left\{\frac{(1-Z)S(Y-m_0(V,X))}{(1-e(V,X))\pi_0(V,X)}\right\}\bigg| Z,S,V,X\right]\right\} \\
&= E\left\{E\left[\,0\,\bigg|Z,S,V,X\right]\right\} = 0,
\end{aligned}$$

so

$$\begin{aligned}
\Delta_\text{C} = E\Bigg[&R_1(V,X)\left\{\frac{ZS(Y-m_1(V,X))}{e(V,X)\pi_1(V,X)} + m_1(V,X) - m_0(V,X)\right\} \\
&- R_0(V,X)\left\{\frac{(1-Z)S(Y-m_0(V,X))}{(1-e(V,X))\pi_0(V,X)}\right\}\Bigg].
\end{aligned} \qquad (3)$$

$\square$

### B.4 A general AIPW type formula with double robustness

Suppose $f(v,x)$ is a function such that $\mu_z(v,x) - f(v,x)$ is irrelevant to $v$, for $z = 0,1,2$. In exactly the same manner as proving Lemma S1 and Theorem 1, we can show that

$$\Delta(x) = E_{V|X=x}\left[R(V,X)\{m_1(V,X) - f(V,X)\}\right] - \{m_0(v,x) - f(v,x)\} \qquad (4)$$

and

$$\Delta_\text{C} = E\left[R_1(V,X)\left\{\frac{ZS(Y-m_1(V,X))}{e(V,X)\pi_1(V,X)} + m_1(V,X) - f(V,X)\right\}\right.$$

9

$$-R_0(V,X)\left\{\frac{(1-Z)S(Y-m_0(V,X))}{(1-e(V,X))\pi_0(V,X)}+m_0(V,X)-f(V,X)\right\}\right]. \quad (5)$$

The choices of $f(v,x)$ are not unique. In fact, $m_0(v,x)$ is a special case of $f(v,x)$.

The models in Equation (5), $\{e(v,x), \pi_z(v,x), m_z(v,x), R_z(v,x), f(v,x) : z = 0, 1\}$, can be estimated parametrically or nonparametrically, resulting in an AIPW type estimator. To see the double robustness of this AIPW type estimator, on the one hand, suppose $m_z(v,x)$ is misspecified as $m_z^*(v,x)$ while other models are correctly specified, then

$$E\left[R_1(V,X)\left\{\frac{ZS(Y-m_1^*(V,X))}{e(V,X)\pi_1(V,X)}+m_1^*(V,X)-f(V,X)\right\}\right]$$
$$=E\left\{E\left[R_1(V,X)\left\{\frac{ZS(Y-m_1^*(V,X))}{e(V,X)\pi_1(V,X)}+m_1^*(V,X)-f(V,X)\right\}\bigg|Z,S,V,X\right]\right\}$$
$$=E\left\{E\left[R_1(V,X)\left\{\frac{ZS(m_1(V,X)-m_1^*(V,X))}{e(V,X)\pi_1(V,X)}+m_1^*(V,X)-f(V,X)\right\}\bigg|V,X\right]\right\}$$
$$=E\left\{E\left[R_1(V,X)\left\{m_1(V,X)-m_1^*(V,X)+m_1^*(V,X)-f(V,X)\right\}\big|V,X\right]\right\}$$
$$=E\left\{E\left[R_1(V,X)\left\{m_1(V,X)-f(V,X)\right\}\big|V,X\right]\right\}$$
$$=E\left[R_1(V,X)\left\{m_1(V,X)-f(V,X)\right\}\right],$$

and similarly,

$$E\left[R_0(V,X)\left\{\frac{(1-Z)S(Y-m_1^*(V,X))}{(1-e(V,X))\pi_0(V,X)}+m_0^*(V,X)-f(V,X)\right\}\right]$$
$$=E\left[R_0(V,X)\left\{m_0(V,X)-f(V,X)\right\}\right],$$

so

$$\Delta_C = E\left[R_1(V,X)\left\{\frac{ZS(Y-m_1^*(V,X))}{e(V,X)\pi_1(V,X)}+m_1^*(V,X)-f(V,X)\right\}\right.$$
$$\left.-R_0(V,X)\left\{\frac{(1-Z)S(Y-m_0^*(V,X))}{(1-e(V,X))\pi_0(V,X)}+m_0^*(V,X)-f(V,X)\right\}\right].$$

On the other hand, suppose $\{e(v,x), \pi_z(v,x)\}$ are misspecified as $\{e^*(v,x), \pi_z^*(v,x)\}$ while other models are correctly specified, then

$$E\left[R_1(V,X)\left\{\frac{ZS(Y-m_1(V,X))}{e^*(V,X)\pi_1^*(V,X)}+m_1(V,X)-f(V,X)\right\}\right]$$
$$=E\left\{E\left[R_1(V,X)\left\{\frac{ZS(Y-m_1(V,X))}{e^*(V,X)\pi_1^*(V,X)}+m_1(V,X)-f(V,X)\right\}\bigg|Z,S,V,X\right]\right\}$$
$$=E\left\{E\left[R_1(V,X)\left\{m_1(V,X)-f(V,X)\right\}\big|V,X\right]\right\}$$
$$=E\left[R_1(V,X)\left\{m_1(V,X)-f(V,X)\right\}\right],$$

and similarly,

$$E\left[R_0(V,X)\left\{\frac{(1-Z)S(Y-m_1(V,X))}{(1-e^*(V,X))\pi_0^*(V,X)}+m_0(V,X)-f(V,X)\right\}\right]$$
$$=E\left[R_0(V,X)\left\{m_0(V,X)-f(V,X)\right\}\right],$$

so

$$\Delta_C = E\left[R_1(V,X)\left\{\frac{ZS(Y-m_1(V,X))}{e^*(V,X)\pi_1^*(V,X)}+m_1(V,X)-f(V,X)\right\}\right.$$

$$-R_0(V,X)\left\{\frac{(1-Z)S(Y-m_0(V,X))}{(1-e^*(V,X))\pi_0^*(V,X)} + m_0(V,X) - f(V,X)\right\}\right].$$

However, it is worth noting that the double robustness has limitations. Since $R_z(v,x)$ has common arguments with $\{e(v,x), \pi_z(v,x)\}$, unfortunately, if $\{e(v,x), \pi_z(v,x)\}$ are misspecified, $R_z(v,x)$ is also likely to be misspecified, which would destroy the consistency of the AIPW type estimator. In practice, the observed survivors in the aggressive treatment group come from two principal strata, so it could be difficult to model $m_1(v,x)$, whereas it is easier to model $m_0(v,x)$. A practical choice is to let $f(v,x) = m_0(v,x)$. When $m_1(v,x)$ is misspecified but other models are correctly specified, the AIPW type estimator is useful.

In a condensed form,

$$\Delta_{\mathrm{C}} = E\left[R_1(V,X)\{m_1(V,X) - f(V,X)\} - R_0(V,X)\{m_0(V,X) - f(V,X)\}\right]. \tag{6}$$

Another doubly robust formula is given by

$$\Delta_{\mathrm{C}} = E\left[R_1(V,X)\left\{\frac{ZS(Y-m_1(V,X))}{e(V,X)\pi_1(V,X)}\right\} - R_0(V,X)\left\{\frac{(1-Z)S(Y-m_0(V,X))}{(1-e(V,X))\pi_0(V,X)}\right\}\right.$$
$$\left.+R_1(V,X)\left\{\frac{(1-Z)S\{m_1(V,X)-m_0(V,X)\}}{(1-e(V,X))\pi_0(V,X)}\right\}\right]. \tag{7}$$

The first two terms have zero means, while the third term has mean

$$E\left[R_1(V,X)\{m_1(V,X) - m_0(V,X)\}\right] = \Delta_{\mathrm{C}}.$$

If either $m_1(v,x)$ is misspecified or $\{e(v,x), \pi_z(v,x)\}$ are misspecified, given other models are correctly specified, Equation (7) still holds. Equation (7) is slightly different from Equation (3), in that Equation (7) uses $R_1(V,X)\{m_1(V,X) - m_0(V,X)\}$ only on the survivors in the $Z=0$ group with weights

$$\frac{(1-Z)S}{(1-e(V,X)\pi_0(V,X)},$$

while Equation (3) uses $R_1(V,X)\{m_1(V,X) - m_0(V,X)\}$ on the full population. Some literature on the doubly robust estimation for average treatment effect on the treated (ATT) under unconfoundedness includes Mercatanti and Li (2014); Shu and Tan (2018).

## C  Proofs of large sample properties

Define the AIPW type estimator

$$\widehat{\Delta}_{\mathrm{C}} = \mathbb{E}_n\left[\widehat{R}_1(V,X)\left\{\frac{ZS(Y-\widehat{m}_1(V,X))}{\widehat{e}(V,X)\widehat{\pi}_1(V,X)} + \widehat{m}_1(V,X) - \widehat{f}(V,X)\right\}\right.$$
$$\left.-\widehat{R}_0(V,X)\left\{\frac{(1-Z)S(Y-\widehat{m}_0(V,X))}{(1-\widehat{e}(V,X))\widehat{\pi}_0(V,X)} + \widehat{m}_0(V,X) - \widehat{f}(V,X)\right\}\right], \tag{8}$$

and an oracle estimator in which all nuisance models are known,

$$\widehat{\Delta}_{\mathrm{C}}^* = \mathbb{E}_n\left[R_1(V,X)\left\{\frac{ZS(Y-m_1(V,X))}{e(V,X)\pi_1(V,X)} + m_1(V,X) - f(V,X)\right\}\right.$$
$$\left.-R_0(V,X)\left\{\frac{(1-Z)S(Y-m_0(V,X))}{(1-e(V,X))\pi_0(V,X)} + m_0(V,X) - f(V,X)\right\}\right], \tag{9}$$

where $\mathbb{E}_n$ means empirical average on the full sample. The oracle estimator $\widehat{\Delta}_{\mathrm{C}}^*$ has good large sample properties because it is an average of $n$ independent random variables,

$$n^{1/2}\left(\widehat{\Delta}_{\mathrm{C}}^* - \Delta_{\mathrm{C}}\right) \xrightarrow{d} N(0, C),$$

where

$$
\begin{aligned}
C &= \mathrm{var}\left[R_1(V,X)\left\{\frac{ZS(Y-m_1(V,X))}{e(V,X)\pi_1(V,X)}+m_1(V,X)-f(V,X)\right\}\right.\\
&\qquad\left. -R_0(V,X)\left\{\frac{(1-Z)S(Y-m_0(V,X))}{(1-e(V,X))\pi_0(V,X)}+m_0(V,X)-f(V,X)\right\}\right]\\
&= E\left[\mathrm{var}\left[R_1(V,X)\left\{\frac{ZS(Y-m_1(V,X))}{e(V,X)\pi_1(V,X)}+m_1(V,X)-f(V,X)\right\}\right.\right.\\
&\qquad\left.\left. -R_0(V,X)\left\{\frac{(1-Z)S(Y-m_0(V,X))}{(1-e(V,X))\pi_0(V,X)}+m_0(V,X)-f(V,X)\right\}\bigg|V,X\right]\right]\\
&\quad+\mathrm{var}\left[E\left[R_1(V,X)\left\{\frac{ZS(Y-m_1(V,X))}{e(V,X)\pi_1(V,X)}+m_1(V,X)-f(V,X)\right\}\right.\right.\\
&\qquad\left.\left. -R_0(V,X)\left\{\frac{(1-Z)S(Y-m_0(V,X))}{(1-e(V,X))\pi_0(V,X)}+m_0(V,X)-f(V,X)\right\}\bigg|V,X\right]\right]\\
&= E\left[\mathrm{var}\left\{\frac{ZSR_1(V,X)(Y-m_1(V,X))}{e(V,X)\pi_1(V,X)}\right\}-\left\{\frac{(1-Z)SR_0(V,X)(Y-m_0(V,X))}{(1-e(V,X))\pi_0(V,X)}\right\}\bigg|V,X\right]\\
&\quad+\mathrm{var}\{\Delta(X)\}\\
&= E\left\{\frac{R_1(V,X)^2\,\mathrm{var}(Y|V,X,Z=1)}{e(V,X)\pi_1(V,X)}-\frac{R_0(V,X)^2\,\mathrm{var}(Y|V,X,Z=0)}{(1-e(V,X))\pi_0(V,X)}\right\}+\mathrm{var}\{\Delta(X)\}.
\end{aligned}
$$

Suppose we use cross-fitting to obtain $\widehat{\Delta}_C$ (Chernozhukov et al., 2018). That is, we randomly divide the full sample into two halves with similar sizes. For the $i$th unit, if it belongs to the $k$th half, then we use the $(3-k)$th half to estimate the models

$$\{e(v,x), \pi_z(v,x), m_z(v,x), R_z(v,x), f(v,x) : z=0,1\},$$

and evaluate them at $i$ to get

$$\{\widehat{e}(v_i,x_i), \widehat{\pi}_z(v_i,x_i), \widehat{m}_z(v_i,x_i), \widehat{R}_z(v_i,x_i), \widehat{f}(v_i,x_i) : z=0,1\},$$

where $i=1,\ldots,n$ and $k=1,2$. Recall that $f(v,x)$ is any function such that $\mu_z(v,x)-f(v,x)$ is irrelevant to $v$, whose simplest choice is $f(v,x)=m_0(v,x)$. Suppose all models are sup-norm consistent,

$$\sup_{v,x}|\widehat{e}(v,x)-e(v,x)|\xrightarrow{p}0,\quad \sup_{v,x}|\widehat{\pi}_z(v,x)-\pi_z(v,x)|\xrightarrow{p}0,\quad \sup_{v,x}|\widehat{R}_z(v,x)-R_z(v,x)|\xrightarrow{p}0,$$

$$\sup_{v,x}|\widehat{m}_z(v,x)-m_z(v,x)|\xrightarrow{p}0,\quad \sup_{v,x}|\widehat{f}(v,x)-f(v,x)|\xrightarrow{p}0,$$

for $z=0,1$. Further assume risk decay as follows, which is implied by the fact that all models converge at rates faster than $o_p(n^{-1/4})$,

$$E\big[\{\widehat{e}(V,X)\widehat{\pi}_1(V,X)-e(V,X)\pi_1(V,X)\}^2\big]\cdot E\big[\widehat{R}_1(V,X)^2\{\widehat{m}_1(V,X)-m_1(V,X)\}^2\big]\to o(n^{-1}),$$
$$E\big[\{(1-\widehat{e}(V,X))\widehat{\pi}_1(V,X)-(1-e(V,X))\pi_1(v,x)\}^2\big]\cdot E\big[\widehat{R}_0(V,X)^2\{\widehat{m}_0(V,X)-m_0(V,X)\}^2\big]\to o(n^{-1}),$$
$$E\big[\{\widehat{R}_z(V,X)-R_z(V,X)\}^2\big]\cdot E\big[\{\widehat{f}(V,X)-f(V,X)\}^2\big]\to o(n^{-1}),$$

for $z=0,1$.

**Proof of Theorem 2.** The AIPW type estimator can be rewritten as

$$\widehat{\Delta}_C = \frac{|\mathcal{I}_1|}{n}\widehat{\Delta}_C^{(1)} + \frac{|\mathcal{I}_2|}{n}\widehat{\Delta}_C^{(2)},$$

and the oracle estimator can be rewritten as

$$\widehat{\Delta}_{\mathrm{C}}^* = \frac{|\mathcal{I}_1|}{n}\widehat{\Delta}_{\mathrm{C}}^{(1)*} + \frac{|\mathcal{I}_2|}{n}\widehat{\Delta}_{\mathrm{C}}^{(2)*},$$

with

$$\widehat{\Delta}_{\mathrm{C}}^{(k)} = \mathbb{E}_{\mathcal{I}_k}\left[\widehat{R}_1(V,X)\left\{\frac{ZS(Y-\widehat{m}_1(V,X))}{\widehat{e}(V,X)\widehat{\pi}_1(V,X)} + \widehat{m}_1(V,X) - \widehat{f}(V,X)\right\}\right.$$
$$\left. - \widehat{R}_0(V,X)\left\{\frac{(1-Z)S(Y-\widehat{m}_0(V,X))}{(1-\widehat{e}(V,X))\widehat{\pi}_0(V,X)} + \widehat{m}_0(V,X) - \widehat{f}(V,X)\right\}\right],$$

$$\widehat{\Delta}_{\mathrm{C}}^{(k)*} = \mathbb{E}_{\mathcal{I}_k}\left[R_1(V,X)\left\{\frac{ZS(Y-m_1(V,X))}{e(V,X)\pi_1(V,X)} + m_1(V,X) - f(V,X)\right\}\right.$$
$$\left. - R_0(V,X)\left\{\frac{(1-Z)S(Y-m_0(V,X))}{(1-e(V,X))\pi_0(V,X)} + m_0(V,X) - f(V,X)\right\}\right],$$

where $\mathcal{I}_k$ is the $k$th half of the sample and $|\mathcal{I}_k|$ is its size. The empirical average is taken upon $\mathcal{I}_k$. It can be easily seen that

$$\widehat{\Delta}_{\mathrm{C}} - \widehat{\Delta}_{\mathrm{C}}^* = \frac{|\mathcal{I}_1|}{n}\left(\widehat{\Delta}_{\mathrm{C}}^{(1)} - \widehat{\Delta}_{\mathrm{C}}^{(1)*}\right) + \frac{|\mathcal{I}_2|}{n}\left(\widehat{\Delta}_{\mathrm{C}}^{(2)} - \widehat{\Delta}_{\mathrm{C}}^{(2)*}\right).$$

Define an intermediate estimator

$$\widehat{\Delta}_{\mathrm{C}}^{(k)\star} = \mathbb{E}_{\mathcal{I}_k}\left[\widehat{R}_1(V,X)\left\{\frac{ZS(Y-m_1(V,X))}{e(V,X)\pi_1(V,X)} + m_1(V,X)\right\}\right.$$
$$- \widehat{R}_0(V,X)\left\{\frac{(1-Z)S(Y-m_0(V,X))}{(1-e(V,X))\pi_0(V,X)} + m_0(V,X)\right\}$$
$$\left. - \left\{\widehat{R}_1(V,X) - \widehat{R}_0(V,X)\right\}\widehat{f}(V,X)\right],$$

and

$$\widehat{\Delta}_{\mathrm{C}}^{\star} = \frac{|\mathcal{I}_1|}{n}\widehat{\Delta}_{\mathrm{C}}^{(1)\star} + \frac{|\mathcal{I}_2|}{n}\widehat{\Delta}_{\mathrm{C}}^{(2)\star}.$$

For notation brevity, we omit the random variables arguments of the models. We first show that $\widehat{\Delta}_{\mathrm{C}}^{(k)}$ is root-$n$ equivalent to $\widehat{\Delta}_{\mathrm{C}}^{(k)\star}$.

$$\widehat{\Delta}_{\mathrm{C}}^{(k)} - \widehat{\Delta}_{\mathrm{C}}^{(k)\star}$$
$$= \mathbb{E}_{\mathcal{I}_k}\left[\widehat{R}_1(\widehat{m}_1 - m_1)\left(1 - \frac{ZS}{e\pi_1}\right) - \widehat{R}_0(\widehat{m}_0 - m_0)\left(1 - \frac{(1-Z)S}{(1-e)\pi_0}\right)\right.$$
$$+ ZS\widehat{R}_1(Y-m_1)\left(\frac{1}{\widehat{e}\widehat{\pi}_1} - \frac{1}{e\pi_1}\right) - (1-Z)S\widehat{R}_0(Y-m_0)\left(\frac{1}{(1-\widehat{e})\widehat{\pi}_0} - \frac{1}{(1-e)\pi_0}\right)$$
$$\left. - ZS\widehat{R}_1(\widehat{m}_1-m_1)\left(\frac{1}{\widehat{e}\widehat{\pi}_1} - \frac{1}{e\pi_1}\right) + (1-Z)S\widehat{R}_0(\widehat{m}_0-m_0)\left(\frac{1}{(1-\widehat{e})\widehat{\pi}_0} - \frac{1}{(1-e)\pi_0}\right)\right].$$

Note that conditioning on $\mathcal{I}_{3-k}$, the estimated models would be fixed. By the positivity assumption, there is a constant $0 < c < 1$ so that $c < e\pi_1 < 1-c$ and $c < (1-e)\pi_0 < 1-c$. For the first terms,

$$E\left[\mathbb{E}_{\mathcal{I}_k}\left\{\widehat{R}_1(\widehat{m}_1-m_1)\left(1-\frac{ZS}{e\pi_1}\right)\right\}\right]^2$$
$$= E\left[E\left\{\left[\mathbb{E}_{\mathcal{I}_k}\left\{\widehat{R}_1(\widehat{m}_1-m_1)\left(1-\frac{ZS}{e\pi_1}\right)\right\}\right]^2\Big|\mathcal{I}_{3-k}\right\}\right]$$

$$
\begin{aligned}
&= E\left[\operatorname{var}\left[\mathbb{E}_{\mathcal{I}_k}\left\{\widehat{R}_1(\widehat{m}_1 - m_1)\left(1 - \frac{ZS}{e\pi_1}\right)\right\}\bigg|\mathcal{I}_{3-k}\right]\right] \\
&= \frac{1}{\mathcal{I}_k} E\left[\operatorname{var}\left\{\widehat{R}_1(\widehat{m}_1 - m_1)\left(1 - \frac{ZS}{e\pi_1}\right)\bigg|\mathcal{I}_{3-k}\right\}\right] \\
&= \frac{1}{\mathcal{I}_k} E\left[E\left\{\widehat{R}_1^2(\widehat{m}_1 - m_1)^2\left(\frac{1}{e\pi_1} - 1\right)\bigg|\mathcal{I}_{3-k}\right\}\right] \\
&\leq \frac{1}{c|\mathcal{I}_k|} E\left\{\widehat{R}_1^2(\widehat{m}_1 - m_1)^2\right\} \\
&= o(n^{-1}).
\end{aligned}
$$

Similarly,

$$
E\left[\mathbb{E}_{\mathcal{I}_k}\left\{\widehat{R}_0(\widehat{m}_0 - m_0)\left(1 - \frac{(1-Z)S}{(1-e)\pi_0}\right)\right\}\right]^2 = o(n^{-1}).
$$

For the second terms, in a similar way, we can show that

$$
E\left[\mathbb{E}_{\mathcal{I}_k}\left\{ZS\widehat{R}_1(Y - m_1)\left(\frac{1}{\widehat{e}\widehat{\pi}_1} - \frac{1}{e\pi_1}\right)\right\}\right]^2 = o(n^{-1}),
$$

$$
E\left[\mathbb{E}_{\mathcal{I}_k}\left\{(1-Z)S\widehat{R}_0(Y - m_0)\left(\frac{1}{(1-\widehat{e})\widehat{\pi}_0} - \frac{1}{(1-e)\pi_0}\right)\right\}\right]^2 = o(n^{-1}).
$$

For the third terms, use the risk decay assumption and Cauchy-Schwarz inequality,

$$
\begin{aligned}
&\left[\mathbb{E}_{\mathcal{I}_k}\left\{ZS\widehat{R}_1(\widehat{m}_1 - m_1)\left(\frac{1}{\widehat{e}\widehat{\pi}_1} - \frac{1}{e\pi_1}\right)\right\}\right]^2 \\
&\leq \mathbb{E}_{\mathcal{I}_k}\left\{ZS\widehat{R}_1^2(\widehat{m}_1 - m_1)^2\right\}\mathbb{E}_{\mathcal{I}_k}\left\{\left(\frac{1}{\widehat{e}\widehat{\pi}_1} - \frac{1}{e\pi_1}\right)^2\right\} \\
&\leq \frac{1}{c^4}\mathbb{E}_{\mathcal{I}_k}\left\{\widehat{R}_1^2(\widehat{m}_1 - m_1)^2\right\}\mathbb{E}_{\mathcal{I}_k}\left\{(\widehat{e}\widehat{\pi}_1 - e\pi_1)^2\right\} \\
&= o_p(n^{-1}).
\end{aligned}
$$

Similarly,

$$
\left[\mathbb{E}_{\mathcal{I}_k}\left\{(1-Z)S\widehat{R}_0(\widehat{m}_0 - m_0)\left(\frac{1}{(1-\widehat{e})\widehat{\pi}_0} - \frac{1}{(1-e)\pi_0}\right)\right\}\right]^2 = o_p(n^{-1}).
$$

Therefore,

$$
\left(\widehat{\Delta}_{\mathrm{C}}^{(k)} - \widehat{\Delta}_{\mathrm{C}}^{(k)\star}\right)^2 = o_p(n^{-1}).
$$

That is, $n^{1/2}\left(\widehat{\Delta}_{\mathrm{C}}^{(k)} - \widehat{\Delta}_{\mathrm{C}}^{(k)\star}\right) \xrightarrow{p} 0$. The discrepancy of $\widehat{\Delta}_{\mathrm{C}}^{(k)\star}$ from $\widehat{\Delta}_{\mathrm{C}}^{(k)*}$ is

$$
\begin{aligned}
&\widehat{\Delta}_{\mathrm{C}}^{(k)\star} - \widehat{\Delta}_{\mathrm{C}}^{(k)*} \\
&= \mathbb{E}_{\mathcal{I}_k}\left[(\widehat{R}_1 - R_1)\left\{\frac{ZS(Y - m_1)}{e\pi_1} + m_1\right\} - (\widehat{R}_0 - R_0)\left\{\frac{(1-Z)S(Y - m_0)}{(1-e)\pi_0} + m_0\right\}\right.\\
&\qquad\left. - \left\{(\widehat{R}_1 - \widehat{R}_0)\widehat{f} - (R_1 - R_0)f\right\}\right] \\
&= \mathbb{E}_{\mathcal{I}_k}\left[(\widehat{R}_1 - R_1)m_1 - (\widehat{R}_0 - R_0)m_0 - \left\{(\widehat{R}_1 - \widehat{R}_0)\widehat{f} - (R_1 - R_0)f\right\}\right] + o_p(n^{-1/2}) \\
&= \mathbb{E}_{\mathcal{I}_k}\left[(\widehat{R}_1 - R_1)m_1 - (\widehat{R}_0 - R_0)m_0 - (\widehat{R}_1 - R_1 - \widehat{R}_0 + R_0)(\widehat{f} - f)\right.
\end{aligned}
$$

14

$$-(R_1 - R_0)(\widehat{f} - f) - (\widehat{R}_1 - R_1 - \widehat{R}_0 + R_0)f\Big] + o_p(n^{-1/2})$$
$$= \mathbb{E}_{\mathcal{I}_k}\Big[(\widehat{R}_1 - R_1)m_1 - (\widehat{R}_0 - R_0)m_0 - (R_1 - R_0)(\widehat{f} - f) - (\widehat{R}_1 - R_1 - \widehat{R}_0 + R_0)f\Big] + o_p(n^{-1/2}),$$

because
$$\mathbb{E}_{\mathcal{I}_k}\left[(\widehat{R}_1 - R_1)\left\{\frac{ZS(Y - m_1)}{e\pi_1}\right\}\right] = \mathbb{E}_{\mathcal{I}_k}\left[(\widehat{R}_0 - R_0)\left\{\frac{(1-Z)S(Y - m_0)}{(1-e)\pi_0}\right\}\right] = o_p(n^{-1/2}),$$
$$\mathbb{E}_{\mathcal{I}_k}\left[(\widehat{R}_1 - R_1 - \widehat{R}_0 + R_0)(\widehat{f} - f)\right] = o_p(n^{-1/2}).$$

Define a drift term
$$D_n = \mathbb{E}_n\left\{(\widehat{R}_1 - R_1)m_1 - (\widehat{R}_0 - R_0)m_0 - (R_1 - R_0)(\widehat{f} - f) - (\widehat{R}_1 - R_1 - \widehat{R}_0 + R_0)f\right\},$$

then
$$n^{1/2}\left(\widehat{\Delta}_C^\star - \widehat{\Delta}_C^* - D_n\right) \xrightarrow{p} 0.$$

By the law of $\widehat{\Delta}_C^*$,
$$n^{1/2}\left(\widehat{\Delta}_C^\star - \Delta_C - D_n\right) \xrightarrow{d} N(0, C).$$

Since $D_n \xrightarrow{p} 0$ by the sup-norm convergence, we have $\widehat{\Delta}_C \xrightarrow{p} \Delta_C$. □

If we let $f(v, x) = m_0(v, x)$, then the drift term $D_n$ can be further simplified as
$$D_n = \mathbb{E}_n\left\{(\widehat{R}_1 - R_1)m_1 - (\widehat{R}_0 - R_0)m_0 - (R_1 - R_0)(\widehat{m}_0 - m_0) - (\widehat{R}_1 - R_1 - \widehat{R}_0 + R_0)m_0\right\}$$
$$= \mathbb{E}_n\left\{(\widehat{R}_1 - R_1)(m_1 - m_0) - (R_1 - R_0)(\widehat{m}_0 - m_0)\right\}.$$

Now suppose the estimators of the parameters in $R_1(v, x; \theta_1)$ and $m_0(v, x; \theta_0)$ are regular and asymptotically linear (RAL), with influence functions $\psi_1(O)$ and $\psi_0(O)$ respectively,
$$\widehat{\theta}_1 - \theta_1 = \mathbb{E}_n\{\psi_1(O)\}, \quad \widehat{\theta}_0 - \theta_0 = \mathbb{E}_n\{\psi_0(O)\}.$$

Then
$$n^{1/2}D_n = n^{1/2}\mathbb{E}_n\left\{\frac{\partial R_1}{\partial \theta_1}(\widehat{\theta}_1 - \theta_1)(m_1 - m_0) + o_p(\widehat{\theta}_1 - \theta_1) - \frac{\partial m_0}{\partial \theta_0}(\widehat{\theta}_0 - \theta_0)(R_1 - R_0) - o_p(\widehat{\theta}_0 - \theta_0)\right\}$$
$$= \mathbb{E}_n\left\{\frac{\partial R_1}{\partial \theta_1}(m_1 - m_0)\right\}n^{1/2}(\widehat{\theta}_1 - \theta_1) - \mathbb{E}_n\left\{\frac{\partial m_0}{\partial \theta_0}(R_1 - R_0)\right\}n^{1/2}(\widehat{\theta}_0 - \theta_0) + o_p(1)$$
$$= E\left\{\frac{\partial R_1}{\partial \theta_1}(m_1 - m_0)\right\}n^{1/2}(\widehat{\theta}_1 - \theta_1) - E\left\{\frac{\partial m_0}{\partial \theta_0}(R_1 - R_0)\right\}n^{1/2}(\widehat{\theta}_0 - \theta_0) + o_p(1)$$
$$= n^{1/2}\mathbb{E}_n\left\{\Gamma_1\psi_1(O) - \Gamma_0\psi_0(O)\right\} + o_p(1).$$

Classical regularity conditions should make the equations above hold. Let
$$\phi(O) = R_1\left(\frac{ZS(Y - m_1)}{e\pi_1} + m_1 - m_0\right) - R_0\left(\frac{(1-Z)S(Y - m_0)}{(1-e)\pi_0}\right),$$

so $\widehat{\Delta}_C^* - \Delta_C = \mathbb{E}_n\{\phi(O) - \Delta_C\}$. Then it follows that
$$n^{1/2}\left(\widehat{\Delta}_C - \Delta_C\right) = n^{1/2}\mathbb{E}_n\left\{\phi(O) - \Delta_C + \Gamma_1\psi_1(O) - \Gamma_1\psi_0(O)\right\} + o_p(1),$$

and hence $\widehat{\Delta}_C$ is regular and asymptotic linear with influence fuction
$$\psi(O) = \phi(O) - \Delta_C + \Gamma_1\psi_1(O) - \Gamma_0\psi_0(O).$$

Take care that $\Gamma_z$ is a row vector and $\psi_z(O)$ is a column vector if $\theta_z$ is a vector ($z = 0, 1$). An asymptotic law of $\widehat{\Delta}_C$ is given by
$$n^{1/2}\left(\widehat{\Delta}_C - \Delta_C\right) \xrightarrow{d} N\left(0, E\{\psi(O)^2\}\right).$$

15

# D    Relaxing nondifferential substitution by external data

With external interventional data, we can determine the value of $\rho(v, x)$. Suppose some individuals in the potential target population eligible for the aggressive treatment were drawn and randomized to receive either the aggressive or conservative treatment, resulting in a new set of interventional data $\{R_i = (Z_i, S_i, X_i, V_i) : i = n+1, \ldots, m\}$. The outcome $Y$ may be time consuming to collect and it may exceed the period of a randomized trial to collect $Y$, so $Y$ may not be included in the interventional data (Kallus and Mao, 2020). This setting is reasonable for post-marketing safety or effectiveness assessment, after a drug has proven beneficial for survival in some previous randomized trials.

We use the notation $R = 1$ to denote that the individual belongs to the interventional data ($R_i = 1$ for $n+1 \leq i \leq m$). The random draw satisfies $R \perp\!\!\!\perp G \mid X, V, Z = 1$. In the interventional data, we can naturally assume $S$-ignorability $S(0) \perp\!\!\!\perp Z \mid X, V, R = 1$. Thus,

$$
\begin{aligned}
\pi_{\text{DD}}(v, x) &= \text{pr}(G = \text{DD} \mid Z = 1, V = v, X = x) \\
&= \text{pr}(G = \text{DD} \mid Z = 1, V = v, X = x, R = 1) \\
&= \text{pr}(S(1) = 0 \mid Z = 1, V = v, X = x, R = 1) \\
&= \text{pr}(S = 0 \mid Z = 1, V = v, X = x, R = 1), \\
\pi_{\text{LD}}(v, x) &= \text{pr}(G = \text{LD} \mid Z = 1, V = v, X = x) \\
&= \text{pr}(G = \text{LD} \mid Z = 1, V = v, X = x, R = 1) \\
&= \text{pr}(S(1) = 1, S(0) = 0 \mid Z = 1, V = v, X = x, R = 1) \\
&= \text{pr}(S(1) = 1 \mid Z = 1, V = v, X = x, R = 1) \\
&\quad - \text{pr}(S(1) = 1, S(0) = 1 \mid Z = 1, V = v, X = x, R = 1) \\
&= \text{pr}(S(1) = 1 \mid Z = 1, V = v, X = x, R = 1) - \text{pr}(S(0) = 1 \mid Z = 1, V = v, X = x, R = 1) \\
&= \text{pr}(S(1) = 1 \mid Z = 1, V = v, X = x, R = 1) - \text{pr}(S(0) = 1 \mid Z = 0, V = v, X = x, R = 1) \\
&= \text{pr}(S = 1 \mid Z = 1, V = v, X = x, R = 1) - \text{pr}(S = 1 \mid Z = 0, V = v, X = x, R = 1).
\end{aligned}
$$

Therefore, we can express $\rho(v, x)$ as

$$
\begin{aligned}
\rho(v, x) &= \frac{\pi_{\text{DD}}(v, x)}{\pi_{\text{DD}}(v_0, x)} \frac{\pi_{\text{LD}}(v_0, x)}{\pi_{\text{LD}}(v, x)} \\
&= \frac{\text{pr}(S = 0 \mid Z = 1, V = v, X = x, R = 1)}{\text{pr}(S = 0 \mid Z = 1, V = v_0, X = x, R = 1)} \\
&\quad \cdot \frac{\text{pr}(S = 1 \mid Z = 1, V = v_0, X = x, R = 1) - \text{pr}(S = 1 \mid Z = 0, V = v_0, X = x, R = 1)}{\text{pr}(S = 1 \mid Z = 1, V = v, X = x, R = 1) - \text{pr}(S = 1 \mid Z = 0, V = v, X = x, R = 1)}
\end{aligned}
$$

or

$$
\begin{aligned}
\rho(v, x) &= \frac{\text{pr}(S = 0 \mid Z = 1, V = v, X = x)}{\text{pr}(S = 0 \mid Z = 1, V = v_0, X = x)} \\
&\quad \cdot \frac{\text{pr}(S = 1 \mid Z = 1, V = v_0, X = x) - \text{pr}(S = 1 \mid Z = 0, V = v_0, X = x, R = 1)}{\text{pr}(S = 1 \mid Z = 1, V = v, X = x) - \text{pr}(S = 1 \mid Z = 0, V = 1, X = x, R = 1)}.
\end{aligned}
$$

Since the proportions of all principal strata in the aggressive treatment group can be identified from the interventional data, apart from the survivor average causal effect on the conservative treatment $\Delta_{\text{C}}$, we can identify the survivor average causal effect on the aggressive treatment $\Delta_{\text{T}}$ and on the whole population $\Delta$.

# E More simulation results

## E.1 Boxplots of bias by different esimation methods

Figure S3 shows the bias of different methods to estimate the SACEC. In Setting 1, the principal strata confound the treatment–survival–outcome processes, without exclusion restriction. In Setting 2, the principal strata confound the survival–outcome processes, with exclusion restriction. In Setting 3, the principal strata confound the treatment–survival processes, with exclusion restriction. In Setting 4, the principal strata confound the treatment–survival–outcome processes, with heterogeneous treatment effects and exclusion restriction.

In all these four settings, the proposed AIPW type estimator is asymptotically unbiased. Since the proposed estimator does not impose any restriction on treatment–survival or survival–outcome processes, the proposed method has large variance especially in Setting 2 where there is actually no unmeasured confounding between $Z$ and $S$. Due to model misspecification, the regression estimator is often biased.

## E.2 Influence of link functions

We have metioned that different choices of the link function $h(v, x)$ can lead to a unique solution. Now we compare three link functions:

$$h_1(v, x) = v - E(V|X = x),$$
$$h_2(v, x) = \pi_1(v, x) - E\{\pi_1(V, X) \mid X = x\},$$
$$h_3(v, x) = \pi_1(v, x)^{-1} - E\{\pi_1(V, X)^{-1} \mid X = x\}.$$

The first link function has the simplest form and is easy for computation. If there are multiple substitutional variabes, i.e., $V$ is a vector $V = (V^1, \ldots, V^q)$, we can take

$$h_1(v, x) = \prod_{j=1}^{q} \left\{ v^j - E(V^j|X = x) \right\}.$$

We use these three links to construct the AIPW type estimator. The resulting boxplots of bias are presented in Figure S4. The difference among these link functions is insignificant.

## E.3 Assessment of confidence intervals

In this section, we empirically examine the influence of the drift term $D_n$ on the inference of the estimation of SACEC. The asymptotic variance $C$ of the oracle estimator $\widehat{\Delta}_C$ can be consistently estimated from data. Due to the drift term $D_n$, the interval

$$\left[ \widehat{\Delta}_C - z_{1-\alpha/2}\widehat{C}, \ \widehat{\Delta}_C + z_{1-\alpha/2}\widehat{C} \right]$$

may not yield approximate $(1 - \alpha)100\%$ coverage of $\Delta_C$ in repeated samplings, where $z_{1-\alpha/2}$ is the upper $(1 - \alpha/2)$ quantile of the standard normal distribution. We call this interval the pseudo confidence interval.

Note that the influence function of $\widehat{\Delta}_C$ can be derived if we use a simple parametric model. We use a linear link $h(v, x) = v - E(V|X = x)$ so that

$$R_1(v, x) = \frac{\exp(v\theta_1)\{v - E(V|X = x)\}}{E\left[\exp(V\theta_1)\{V - E(V|X = x) \mid X = x\}\right]} \cdot \frac{\text{pr}(Z = 0, S = 1 \mid X = x)}{\text{pr}(Z = 0, S = 1)}$$

and a linear outcome regression model $m_0(v, x) = (1, v, x)\theta_0$. The likelihoods involving $\theta_1$ and $\theta_0$ are separable, so $\widehat{\theta}_1$ and $\widehat{\theta}_0$ are independent. Plug the estimated parameters in $\psi(O)$ and get the estimated influence function $\widehat{\psi}(O)$, or directly calcute a conservative version

$$\mathbb{E}_n\{\widehat{\psi}(O)^2\} \approx \widehat{C} + \widehat{\Gamma}_1\widehat{\text{var}}(\widehat{\theta}_1)\widehat{\Gamma}_1^{\mathrm{T}} + \widehat{\Gamma}_0\widehat{\text{var}}(\widehat{\theta}_0)\widehat{\Gamma}_0^{\mathrm{T}},$$

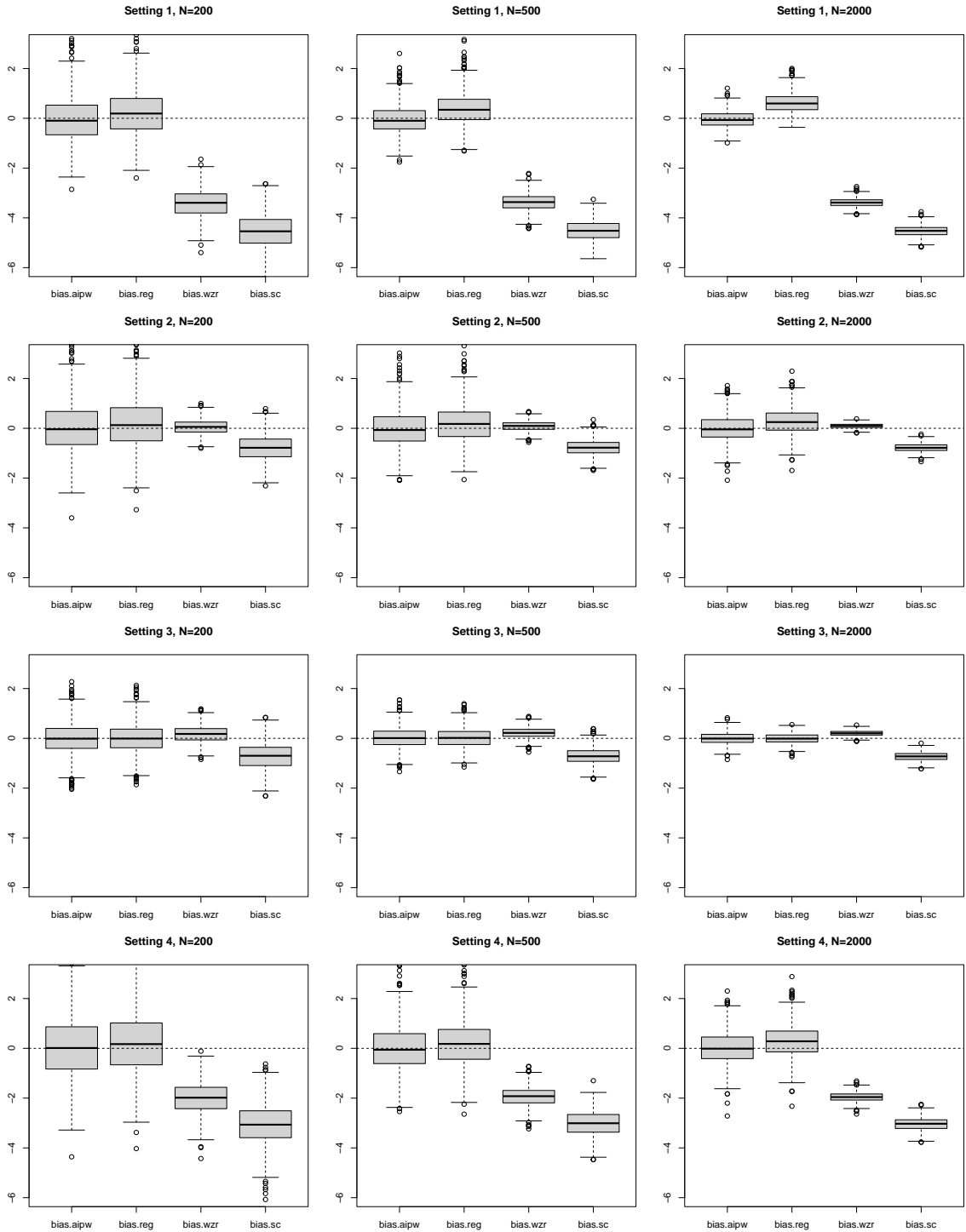

Figure S3: Boxplots of bias in estimating the SACEC, by survivor-case analysis (SC), Wang et al. (2017)'s method (WZR) and proposed method including regression estimator (REG) and AIPW type estimator (AIPW).

18

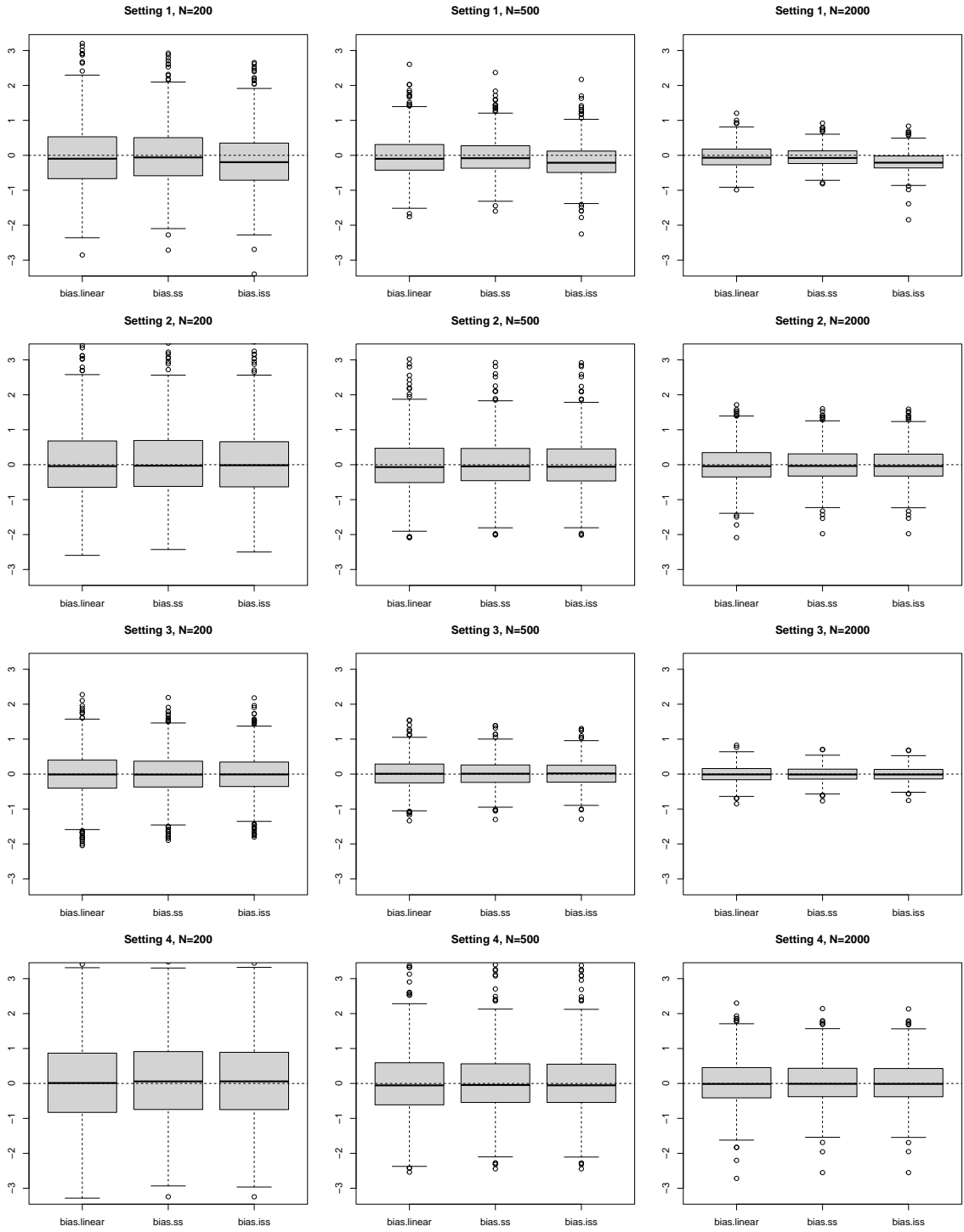

Figure S4: Boxplots of bias in estimating the SACEC, using different links $h_1$ (linear), $h_2$ (ss) and $h_3$ (iss).

Table S2: Coverage and width of the 95% confidence intervals associated with the AIPW type estimator using the pseudo confidence interval (PCI), asymptotic confidence interval (ACI) and bootstrap confidence interval (BCI)

| Setting | $n$ | Coverage | | | Width | | |
|---|---|---|---|---|---|---|---|
| | | PCI | ACI | BCI | PCI | ACI | BCI |
| 1 | 200 | 0.973 | 0.983 | 0.955 | 4.986 | 5.357 | 3.522 |
| | 500 | 0.986 | 0.992 | 0.947 | 3.834 | 4.058 | 2.297 |
| | 2000 | 1.000 | 1.000 | 0.946 | 2.566 | 2.671 | 1.290 |
| 2 | 200 | 0.984 | 0.988 | 0.947 | 6.132 | 7.092 | 4.097 |
| | 500 | 0.990 | 0.992 | 0.948 | 4.982 | 5.184 | 2.956 |
| | 2000 | 0.993 | 0.994 | 0.952 | 3.854 | 3.940 | 2.057 |
| 3 | 200 | 0.761 | 0.972 | 0.947 | 1.345 | 2.316 | 2.434 |
| | 500 | 0.732 | 0.969 | 0.944 | 0.910 | 1.575 | 1.614 |
| | 2000 | 0.758 | 0.961 | 0.952 | 0.550 | 0.895 | 0.921 |
| 4 | 200 | 0.991 | 0.993 | 0.957 | 6.759 | 7.043 | 4.215 |
| | 500 | 0.994 | 0.997 | 0.951 | 5.164 | 5.335 | 2.769 |
| | 2000 | 1.000 | 1.000 | 0.955 | 3.471 | 3.551 | 1.560 |

where $\widehat{C}$ and $\widehat{\Gamma}_z$ are the empirical counterparts of $C$ and $\Gamma_z$ ($z = 0, 1$). According to the large sample property of $\widehat{\Delta}_C$, we can construct an asymptotic $(1 - \alpha)100\%$ (conservative) confidence interval as

$$\left[\widehat{\Delta}_C - z_{1-\alpha/2}\mathbb{E}_n\{\widehat{\psi}(O)^2\}, \ \widehat{\Delta}_C + z_{1-\alpha/2}\mathbb{E}_n\{\widehat{\psi}(O)^2\}\right].$$

We simulate 1000 times and calculate the average coverage and width of this pseudo confidence interval (PCI), conservative asymptotic confidence interval (ACI) and bootstrap confidence interval (BCI) at the nominal level $\alpha = 0.05$ associated with the AIPW type estimator (8). The results are listed in Table S2. The PCI is conservative in Settings 1, 2 and 4 (where $m_1(v, x)$ is misspecified) but anti-conservative in Setting 3 (where $m_1(v, x)$ is correctly specified). Generally the BCI has a coverage close to the nominal level, no matter whether $m_1(v, x)$ is correctly specified or not. In Setting 3, when $m_1(v, x)$ is correctly specified, the ACI has a shorter width but higher coverage than BCI. When the user has confidence that models can be correctly specfied, ACI is recommended, which has smaller computation cost than bootstrap. Otherwise ACI is a conservative confidence interval. However, in practice, to avoid the consequence of model misspecification, BCI is the safest.

### E.4 Sensitivity analysis for monotonicity

Suppose that in the simulation setting 1 added with exclusion restriction (so that existing methods have smaller bias), we split the LL stratum into two strata: LL and DL. In this case, there will be four principal strata in both treatment groups. To generate the DL stratum, we introduce a sensitivity parameter $\kappa \in (0, 1]$. In the previously generated subsample of $S(1) = S(0) = 1$, we modify $S(1)$ to 0 with probability $1 - \kappa$. When monotonicity holds, $\kappa \equiv 1$.

Figure S5 displays the median bias of the estimator of SACEC in 1000 simulated samples when the sample size $n = 500$. When $\kappa$ is near 1, i.e., monotonicity is just slightly violated, the bias is not large. The proposed AIPW type estimator still outperforms other estimators. However, when $\kappa$ is small, i.e., monotonicity is severely violated, the bias of the AIPW type estimator is large. In fact, violation of monotonicity can destroy the validity of other assumptions such as nondifferential substitution. The target population in the definition of SACEC would not be identifiable any more since the survivors in the conservative treatment group come from two indistinguishable principal strata. Therefore, in practical usage of the proposed method,

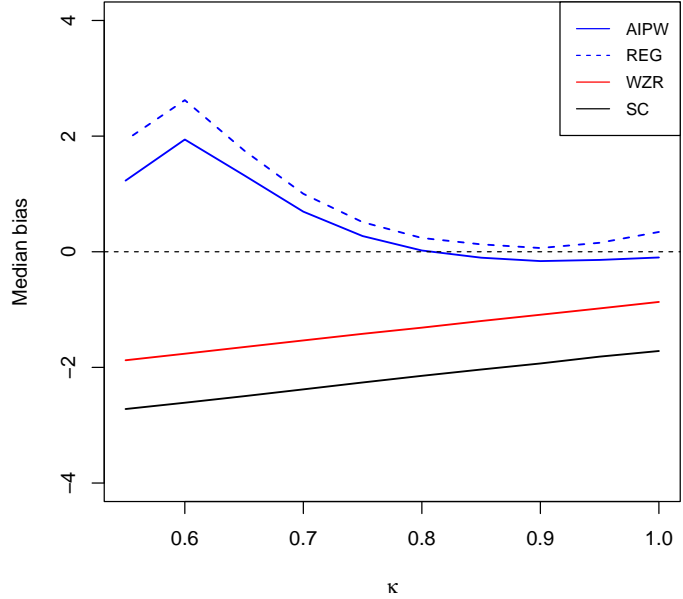

Figure S5: Sensitivity analysis for monotonicity.

practitioners should first try to carefully select a target population in which monotonicity is believed to hold.

## F  Assumptions check in the real data

The first assumption, latent ignorability, is untestable. Latent ignorability is believed to hold because we have controlled all the baseline risk factors that are related with relapse found by clinical experience, namely pre-treatment minimum residual disease (MRD) presence, disease status and diagnosis; see the univariate and multivariate analyses of factors associated with transplantation outcomes in Chang et al. (2020). To control the post-treatment risk factor, chronic graft versus host disease (GVHD), we condition on the principal strata, which is a reflection of potential post-treatment GVHD because acute GVHD can lead to non-relapse mortality (NRM).

Monotonicity can be easily explained be clinical knowledge. To cure leukemia, stem cells were transplanted from a donor to a patient (receiver). MSDT has fewer mismatched human leukocyte antigen (HLA) loci between the donor and the receiver than haplo-SCT, so MSDT would lead to a lower risk of immune rejection (or GVHD) and thus a lower risk of NRM. Regarding monotonicity, this assumption is statistically untestable because $S(1)$ and $S(0)$ can never be jointly observed. We can never know which principal stratum a patient belongs to. However, we know that the age $V$ has a positive effect on $Z$ and negative effect on survival $S$, because older people are more accessible to MSDT ($Z = 1$) but meanwhile are more likely to experience NRM. Since $V$ can be considered as a proxy of the unobserved principal strata $G$, the true effect of $Z$ on $S$ is minified by confounding factors. When regressing $S$ on $Z$, $V$ and $X$, the coefficient of $Z$ is significantly negative. So the true effect of $Z$ on $S$ should be larger than this significant result. This finding does not provide evidence to reject monotonicity.

Next, we comment on the assumptions on the substitutional variable. The rationale that age serves as a substitutional variable has been confirmed by Hangai et al. (2019). Age is a risk factor of NRM but not a risk factor of relapse. To check the substitution relevance, we

consider regressing $S$ on $Z$, $V$ and $X$. The coefficient of $V$ is significantly negative. With the age increasing, the risk of NRM is higher. To check the non-interaction, we must confirm that $V$ has no heterogenous effects on $Y$ between treatment groups. Non-interaction is untestable but can be partially verified. We add an interaction term for $Z$ and $V$ when regressing $Y$ on $Z$, $V$ and $X$ among survivors. The coefficient of the interaction term has an estimate very close to zero, with a $p$-value at 0.348. It is deemed that age should have a similar effect on $Y(1)$ and $Y(0)$ in all survivors because there is no biological mechanism indicating pleiotropy for age between different transplant modalities. Indeed, the estimated coefficient of $V$ is also insignificant. In line with clinical findings, the stronger exclusion restriction that age does not directly influence relapse may hold. Nondifferential substitution is hard to verify, so we performed a sensitivity analysis.



\end{document}